# Why so? or Why no?
# Functional Causality for Explaining Query Answers

Alexandra Meliou    Wolfgang Gatterbauer    Katherine F. Moore    Dan Suciu

Department of Computer Science and Engineering,
University of Washington, Seattle, WA, USA
{ameli,gatter,kfm,suciu}@cs.washington.edu

## ABSTRACT

In this paper, we propose *causality* as a unified framework to *explain query answers* and non-answers, thus generalizing and extending several previously proposed approaches of provenance and missing query result explanations.

We develop our framework starting from the well-studied definition of *actual causes* by Halpern and Pearl [13]. After identifying some undesirable characteristics of the original definition, we propose *functional causes* as a refined definition of causality with several desirable properties. These properties allow us to apply our notion of causality in a database context and apply it uniformly to define the causes of query results and their individual contributions in several ways: (i) we can model both *provenance* as well as *non-answers*, (ii) we can define explanations as either *data* in the input relations or relational *operations* in a query plan, and (iii) we can give graded degrees of responsibility to individual causes, thus allowing us to *rank causes*. In particular, our approach allows us to explain contributions to relational *aggregate functions* and to rank causes according to their respective responsibilities. We give complexity results and describe polynomial algorithms for evaluating causality in tractable cases. Throughout the paper, we illustrate the applicability of our framework with several examples.

Overall, we develop in this paper the theoretical foundations of causality theory in a database context.

## 1. INTRODUCTION

When analyzing data sets and domains of interest, users are often interested in *explanations* for their observations. In a database context, such explanations concern results to explicit or implicit queries. For example, "Why does my personalized newscast have more than 20 items today?" Or, "Why does my favorite undergrad student not appear on the Dean's list this year?" Database research that addresses these or similar questions is mainly work on *lineage of query results*, such as why [8] or where provenance [3], and very recently, explanations for non-answers [17, 4]. While these approaches differ over what the response to questions should be, all of them seem to be linked through a common underlying theme: understanding *causal relationships* in databases.

Humans usually have an intuition about what constitutes a cause of a given effect. In this paper, we define the fundamental notion of *functional causality* that can model this intuition in an exact mathematical framework, and show how it can be applied to encode and solve various causality related problems. In particular, it allows us to uniformly model the questions of WHY SO? and WHY NO? with regards to query answers. It also effectively allows us to represent different approaches taken so far, thus illustrating that causality is a critical element unifying important work in this field.

We start with a simple illustrative example.

EXAMPLE 1.1 (NEWS FEED). *A user has a personalized news feed that filters incoming news based on matching predefined tags. Let relation $K(tag)$ represent the table with the user-defined tags, $N(nid, story, tag)$ the incoming news, and $P(nid, story)$ the personalized news feed. For simplicity, we assume one single tag per news item and ignore timestamps. $P$ can then be represented by the query*

```
create view P
as     select N.nid, N.story
       from   N
       where  exists ( select *
                       from   K
                       where  K.tag=N.tag)
```

*As a result, the view P will be a collection of news matching the user's preferences as shown in Fig. 1. The user may now ask questions about this view. For example, "Why am I getting so many stories about Indianapolis?" (5 in total). The system should answer that the user's keywords* `DB_conf`, `Purdue`, *and* `Movies` *are causes with some kind of decreasing responsibility. On the other hand, the user may have heard that there should be far more news feeds on Indianapolis this week and wonders "Why am I NOT getting MORE stories about Indianapolis?" The system should suggest the lack of the keyword* `Indy_500` *in the user-defined relation K as possible cause (inserting it would increase the count from 5 to 8 articles on Indianapolis).*

As illustrated in Example 1.1, we want to allow users to ask simple questions based on the results they receive, and hence, allow them to learn what may be the *cause* of any surprising or undesirable answer. Such questions can refer

N(ews feeds)

| nid | story | tag |
|---|---|---|
| 1 | ... the race of Indianapolis this year may ... | Indy_500 |
| 2 | ... economic downturn affected sensitive ... | Business |
| 3 | ... with sequences shot in Indianapolis ... | Movies |
| 4 | ... when President Obama meets former ally ... | Obama |
| 5 | ... Indianapolis officials debating the budget ... | Purdue |
| 6 | ... most amazing event in the US with few ... | Burning_man |
| 7 | ... PODS held in Indianapolis this year ... | DB_conf |
| 8 | ... discussed in a recent talk the options to ... | Politics |
| 9 | ... VLDB conference this year in Singapore ... | DB_conf |
| 10 | ... at the the Indianapolis Motor Speedway ... | Indy 500 |
| 11 | ... Indianapolis host to SIGMOD/PODS ... | DB_conf |
| 12 | ... SIGMOD in Indianapolis promises to be ... | DB_conf |
| 13 | ... more people in Indianapolis this year ... | Indy_500 |
| 14 | ... recent ranking held positive surprises for ... | Purdue |

K(eywords)

| tag |
|---|
| Obama |
| DB_conf |
| Purdue |
| Burning_man |
| Movies |
| Afghanistan |

P(ersonalized news)

| nid | story |
|---|---|
| 3 | ... with sequences shot in Indianapolis ... |
| 4 | ... when President Obama meets former ally ... |
| 5 | ... Indianapolis officials debating the budget ... |
| 6 | ... most amazing event in the US with few ... |
| 7 | ... PODS held in Indianapolis this year ... |
| 9 | ... VLDB conference this year in Singapore ... |
| 11 | ... Indianapolis host to SIGMOD/PODS ... |
| 12 | ... SIGMOD in Indianapolis promises to be ... |
| 14 | ... recent ranking held positive surprises for ... |

**Figure 1: Example of a personalized news-feed ($P$) as a result of a query filtering all news ($N$) based on user-defined keywords ($K$).**

to either presence (WHY SO?) or absence (WHY NO?) of results. Furthermore, the user should be provided with a ranking of causes based on their individual contribution or *responsibility*. Our ultimate goal is to define a language that allows users to specify causal queries for given results. In this paper, we lay the theoretical groundwork and define a formal model that allows us to capture such causality-related questions in a uniform framework.

**Summary and outline.** Section 2 analyzes causality in Boolean networks in general: We start by reviewing existing definitions of *counterfactual* and *actual causes* (Sect. 2.1 to 2.2). We also illustrate problems of these previous definitions, and propose *functional causes* as a refined notion of causality that mitigates these problems (Sect. 2.3). In Sect. 3, we then describe and prove several desirable properties of functional causes. We also give complexity results for general and restricted Boolean networks. Section 4 applies our general framework to give WHY SO? and WHY NO? explanations to *database queries*. We show that our unifying framework generalizes provenance as well as non-answers (Sect. 4.1), handles contributions to aggregate functions by ranking causes according to their responsibilities for the result (Sect. 4.2), and can also model causes other than tuples (Sect. 4.3). We discuss related work in Sect. 5, point out some directions for future work (Sect. 6), and give detailed proofs and elaborated examples in the appendix.

## 2. CAUSALITY

This section discusses the two most established notions of causality, then our new definition. The first is the notion of *counterfactual causes*, which is intuitive and simple, but very limited in its applicability. The second is the definition of *actual causes* by Halpern and Pearl (HP from now on), which can better reproduce common-sense causal answers and has become central in the causality literature. We then give our definition of *functional causes* which is a refinement of the HP definition that can model more cases correctly and has additional desirable properties for database applications.

**General notions.** We assume a set of Boolean random variables which model a causal problem. A capital letter (e.g. $X$) denotes a *variable*, and a lower case letter with exponent 0 or 1 (e.g. $x^0$) denotes a *truth value*. An *event* is a truth value assignment to one or more variables (e.g. $X = x^0$). We use the vector sign (e.g. $\vec{X}$) to denote an ordered or unordered set, depending on the context. A *causal model* $M$ is a tuple $(\vec{N}, \mathcal{F})$ with $\vec{N}$ representing a set of variables, and $\mathcal{F} = \{F_N | N \in \vec{N}\}$ a set of *structural equations* $F_N : \{0,1\}^{|\vec{P}_N|} \to \{0,1\}$ that assign a truth value to $N$ for each value of its parents $\vec{P}_N \subseteq \vec{N} \setminus \{N\}$. The *causal network* ($CN$) is the directed acyclic graph representing the dependencies between the Boolean variables (like in a Bayesian network). We call nodes without parents *input variables* and the rest *dependent variables*, denoting them with $\vec{X}$ and $\vec{Y}$, respectively. We associate to each dependent variable $Y$ a *Boolean formula* that determines its truth value $Y(\vec{X})$ based on the values of the input variables. The Boolean formula of a distinguished *effect variable* is denoted as $\Phi(\vec{X})$. The *effect* $\phi$ represents the event that the effect variable has its current assignment $\phi = (\Phi(\vec{X}) = \Phi(\vec{x}^0))$. Causality is always determined for a given *actual assignment* $\vec{x}^0$. The *causal path* is the set of all descendants of a variable under consideration. An *external intervention* $[\vec{S} \leftarrow \vec{s}^1]$ for $\vec{S} \subseteq \vec{N}$ considers a *modified causal model* where each node $N \in \vec{S}$ is assigned a truth value $n^1$ that replaces its structural equation $F_N$.

### 2.1 Counterfactual Causes

With deep roots in philosophy [18], the argument of counterfactual causality is that the relationship between cause and effect can be understood as a counterfactual statement, i.e. an event is considered a cause of an effect if the effect would not have happened in the absence of the event.

DEFINITION 2.1 (COUNTERFACTUAL CAUSE [22]). *The event $X = x^0$ is a cause of $\phi$ in a causal model $M$ iff:*

CC1. $X = x^0 \wedge \phi$
CC2. $[X \leftarrow \neg x^0] \Rightarrow \neg \phi$

EXAMPLE 2.2 (ONE THROWER). *Alice throws a rock at a bottle and the bottle breaks. If Alice had not thrown the rock, then the bottle would not have broken. Therefore, Alice throwing the rock is a cause of the bottle breaking.*

**Shortcomings of Counterfactual Causes.** Counterfactual causality cannot handle slightly more complicated scenarios such as *disjunctive causes*, i.e. when there are two potential causes of an event.

EXAMPLE 2.3 (TWO THROWERS [11]). *Alice and Bob each throw a rock at the bottle and it breaks. Had Alice not thrown the rock, the bottle would still have broken. According to the counterfactual definition, Alice's throw is not a cause even though common sense suggests she should be.*

Figure 2a shows an example of a simple causal network for Example 2.3. The events of Alice and Bob throwing rocks are modeled with truth value 1 for variables $A$ and $B$ respectively, while $Y$ models the effect variable (i.e. the bottle breaking $\phi$) which is true if either $A$ or $B$ is true.

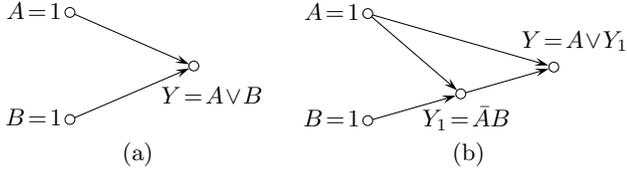

(a)        (b)

**Figure 2: Two throwers.** In both models, the bottle breaks ($Y = 1$) if either Alice throws ($A = 1$) or Bob throws ($B = 1$). Model b encodes the *preemption* of Bob's throw by Alice with an additional intermediate variable $Y_1$ for Bob hitting the unbroken bottle.

## 2.2 Actual Causes

The HP definition of causality [13] is based on counterfactuals, but can correctly model disjunction and many other complications. The idea is that $X$ is a cause of $Y$ if $Y$ counterfactually depends on $X$ under "some" *permissive contingency*, where "some" is elaborately defined. This definition is significant in causality theory. We present it here in an abbreviated way and refer to [13] for details. Note that the HP definition allows *subsets* of both input variables $\vec{X}$ and *dependent variables* $\vec{Y}$ to be a cause of $\phi$. In the following, $\vec{N}_c$ refers to a subset from all nodes of a network.

DEFINITION 2.4 (ACTUAL CAUSE [13], DEF 3.1). *The event $\vec{N}_c(\vec{x}^0) = \vec{n}_c^0$ is a cause of $\phi$ in a causal model $M$ iff:*

AC1. *Both $\vec{N}_c(\vec{x}^0) = \vec{n}_c^0$ and $\phi$ hold under assignment $\vec{x}^0$*

AC2. *There exists a partition $(\vec{Z}, \vec{W})$ of $\vec{N}$ with $\vec{N}_c \subseteq \vec{Z}$ and an assignment $(\vec{n}_c^1, \vec{w}^1)$ of the variables $(\vec{N}_c, \vec{W})$, such that the following two conditions hold:*
  (a) $[\vec{N}_c \leftarrow \vec{n}_i^1, \vec{W} \leftarrow \vec{w}^1] \Rightarrow \neg\phi$
  (b) $[\vec{N}_c \leftarrow \vec{n}_c^0, \vec{W}' \leftarrow \vec{w}'^1, \vec{Z}' \leftarrow \vec{z}'^0] \Rightarrow \phi$, *for all subsets $\vec{W}' \subseteq \vec{W}$ and $\vec{Z}' \subseteq \vec{Z}$*

AC3. *$\vec{N}_c$ is minimal, i.e. no subset $\vec{N}_c' \subset \vec{N}_c$ is a cause.*

The heart of the definition is condition AC2, which is effectively a generalization of counterfactual causes. The requirement is that there exists some assignment of the variables for which $\vec{N}_c$ is counterfactual, and that this assignment does not make any fundamental changes to the *causal path* of $\vec{N}_c$.

The HP definition correctly handles disjunctive causes as in Example 2.3, recognizing both Alice's and Bob's throws as causes. Its use of the causal network makes it very flexible in capturing different scenarios of causal relationships. For example, it is easy to model *preemption*, i.e. when there are two potential causes of an event and one preempts the other.

EXAMPLE 2.5 (TWO THROWERS CONTINUED). *Assume that Alice's rock hits the bottle first. Then Alice's throw would be considered a cause of the bottle breaking, but not Bob's. This precedence of Alice's throw is not encoded in the network of Fig. 2a (model a). It can be modeled by adding the variable $Y_1$ in Fig. 2b (model b): The bottle breaks if either Alice throws, or if Alice doesn't throw and Bob throws ($Y_1 = 1$). The Boolean formulas for the effect, $\Phi(\vec{X}) = A \lor B$ or $\Phi(\vec{X}) = A \lor \bar{A}B$ for models a and b respectively, are equivalent, but the causal relevance of variable $B$ is not: Bob's throw is an actual cause in model a, but no in model b.*

This result is intuitive, because Alice's rock hits the bottle first, breaking it and *preempting* that Bob can hit and break it. While there exists an assignment of variables ($A \leftarrow 0$) that makes Bob's throw ($B = 1$) counterfactual, this assignment changes the value of node $Y_1$ from 0 to 1, establishing a change in the causal path of $B$. Since there is no path from $B$ to $Y$ that doesn't go through $Y_1$, $B$ is not a cause.

**Shortcomings of Actual Causes.** The HP definition of actual cause is well established in the causality literature, but it does not correctly handle some cases, leading to non-intuitive results. The following is a well-studied example (see [22]), originally given by McDermott [21], for which the HP definition does not match common sense, i.e. the commonly accepted interpretation in philosophical circles.

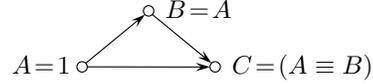

**Figure 3: Shock C.** Simple example where actual causality fails to match common-sense: $A$ is determined a cause of $C$ according to the HP definition, although $C$ is always true.

EXAMPLE 2.6 (SHOCK C [21]). *Shock C is a game for three players. $A$ and $B$ each have a switch which they can move to the left or right. If both switches are thrown into the same position, a third person $C$ receives a shock. $A$ does not want to shock C. Seeing B's switch in the left position, A moves his switch to the right. B wants to shock C. Seeing A's switch thrown to the right, she now moves her switch to the right as well. C receives a shock. Clearly, A's move was a cause of B's move, and B's move was a cause of C's shock, but A's move was not a cause of C's shock.*

*This example can be modeled with the causal network from Fig. 3 and structural equations*

$$B = A$$
$$C = (A \equiv B) = AB \lor \bar{A}\bar{B}$$

*under actual assignment $A = 1$, and hence $B = 1, C = 1$. The effect $\phi$ under consideration is $C = 1$. Here, and contrary to common sense, $A = 1$ is an actual cause of $C = 1$: Take $\vec{W} = \{B\}$ with $b^1 = 1$. Then AC2(a) holds: $[A \leftarrow 0, B \leftarrow 1] \Rightarrow \neg\phi$. Also AC2(b) holds: $\vec{Z} \setminus \vec{N}_c \setminus \{C\}$ is empty, and for either $\vec{W}' = \vec{W}$ or $\vec{W}' = \emptyset$, $C$ is 1 because of $A \leftarrow 1$. Hence, the HP definition does not deliver the common-sense answer for this example, making A the cause of a tautology.*

Appendix B.1 gives more details on the Shock C example, and shows that the HP definition cannot handle this example even with a more elaborate causal network, while the following definition of functional causality can.

## 2.3 Functional Causes

A fundamental challenge in applying causality to queries is that causality is defined over an entire *network*: it is not enough to know the dependency of the effect on the input variables, we also need to reason about intermediate dependent nodes. This requirement is difficult to carry over to a database setting, where we care about the semantics of a query rather than a particular query plan. Our approach is to represent a causal network with two appropriate functions that *semantically capture* the causal dependencies of a network. The two key notions we need for that are *potential functions* and *dissociation expressions*.

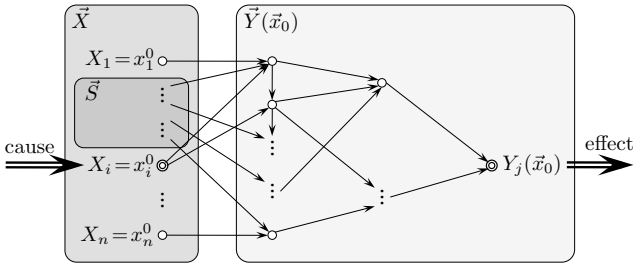

**Figure 4:** FC framework: the causal network is partitioned into the input variables $\vec{X}$ with cause under consideration $X_i$, and dependent variables $\vec{Y}$ with effect variable $Y_j$. Support $\vec{S} \subseteq \vec{X} \setminus \{X_i\}$ corresponds to permissive contingency from the HP framework.

Figure 4 represents a causal network in our framework. In contrast to the HP approach, only input variables from $\vec{X}$ can be causes and part of permissive contingencies. As in the HP approach, every dependent node $Y$ is described by a structural equation $F_Y$, which assigns a truth value to $Y$ based on the values of its parents. The *Boolean formula* $\Phi_Y$ of $Y$ defines its truth assignment based on the input variables $\vec{X}$, and is constructed by recursing through the structural equations of $Y$'s ancestors. For example, in Fig. 2b, $\Phi_Y(\vec{X}) = A \vee (\bar{A} \wedge B)$, where $\vec{X} = \{A, B\}$. We denote as $\Phi(\vec{X}) = \Phi_{Y_j}(\vec{X})$, where $Y_j$ is the effect node, and we say that the causal network has formula $\Phi$. The *potential function* $P_\Phi$ is then simply the unique multilinear polynomial representing $\Phi$. It is equal to the probability that $\Phi$ is true given the probabilities of its input variables.

DEFINITION 2.7 (POTENTIAL FUNCTION). *The potential function $P_\Phi(\vec{x})$ of a Boolean formula $\Phi(\vec{X})$ with probabilities $\vec{x} = \{x_1, \ldots, x_k\}$ of the input variables is defined as follows:*

$$P_\Phi(\vec{x}) = \sum_{\vec{\varepsilon} \to \{0,1\}^k} \left( \prod_{i=1}^k x_i^{\varepsilon_i} \right) \Phi(\vec{\varepsilon}), \quad x_i^{\varepsilon_i} = \begin{cases} x_i & \text{if } \varepsilon_i = 1 \\ 1 - x_i & \text{if } \varepsilon_i = 0 \end{cases}$$

The potential function is a sum with one term for each truth assignment $\vec{\varepsilon}$ of variables $\vec{X}$. Each term is a product of factors of the form $x_i$ or $1 - x_i$ and only occurs in the sum if the formula is true at the given assignment ($\Phi(\vec{\varepsilon}) = 1$). For example, if $\Phi = X_1 \wedge (X_2 \vee X_3)$ then $P_\Phi = x_1 x_2 (1-x_3) + x_1(1-x_2)x_3 + x_1 x_2 x_3$, which simplifies to $x_1(x_2 + x_3 - x_2 x_3)$.

We ground our framework on potential functions because they allow us to extend functional causes to probabilistic databases, a topic that we briefly discuss in Sect. 6. For the deterministic settings of this paper, we use delta notation to denote changes $\Delta P$ in the potential function due to changes in the inputs: Given an actual assignment $\vec{x}^0$ and a subset of variables $\vec{S}$, we define $\Delta P_\Phi(\vec{S}) := P_\Phi(\vec{x}^0) - P_\Phi(\vec{x}^0 \oplus \vec{S})$, where $\vec{x}^0 \oplus \vec{S}$ (denoting XOR) indicates the assignment obtained by starting from $\vec{x}^0$ and inverting all variables in $\vec{S}$.

We use *dissociation expressions (DE)* to semantically capture differences in causality between networks with logically equivalent boolean formulas (e.g. Fig. 2):

DEFINITION 2.8 (DISSOCIATION EXPRESSION). *A dissociation expression with respect to a variable $X_0$ is a Boolean expression defined by the grammar:*

$$\Psi ::= X \in \vec{X}$$
$$\Psi ::= \sigma(\Psi_1, \Psi_2, \ldots, \Psi_k),$$
$$X_0 \in V(\Psi_i) \cup V(\Psi_j) \Rightarrow V(\Psi_i) \cap V(\Psi_j) \subseteq \{X_0\}$$

*where $V(\Psi_i)$ is the set of input variables of formula $\Psi_i$.*

Dissociation expressions allow us to semantically capture with a boolean formula the effect of a variable along different network paths, by disallowing a variable from being combined with $X_0$ in more than one subexpression. For example, in the network of Fig. 5a, variable $A$ contributes to the causal path of $B$ at two locations. This "independent" influence can be represented by the dissociation expression $\Psi = A_1 \vee (\bar{A}_2 \wedge B)$, which essentially separates $A$ into two variables. $\Psi' = A \vee (\bar{A} \wedge B)$ is not a valid DE with respect to $B$, because for its subexpressions, $\Psi'_1 = A$ and $\Psi'_2 = \bar{A} \wedge B$, it is $B \in V(\Psi'_1) \cup V(\Psi'_2)$ but $V(\Psi'_1) \cap V(\Psi'_2) = \{A\} \not\subseteq \{B\}$. Note however that $\Psi'$ is a DE w.r.t $A$, as no variable is combined with $A$ in more than one subexpression.

We demonstrate how $\Psi$ captures semantically the network structure: to check actual causality of $B$ in the network of Fig. 5a, we need to determine the value of $Y$ for a setting $\{A = 0, B = 1\}$ while forcing $Y_1$ to its original value, as part of condition AC2(b). The dissociation expression $\Psi(A_1, A_2, B) = A_1 \vee (\bar{A}_2 \wedge B)$, with potential function $P_\Psi(a_1, a_2, b) = a_1 + b - a_1 b - a_2 b + a_1 a_2 b$, allows us to perform the same check by simply computing $P_\Psi(0, 1, 1)$. In this case $P_\Psi(0, 1, 1) = 0 \neq P_\Psi(1, 1, 1)$, which was the original variable assignment, meaning that the change in assignment altered values on the causal path.

To link dissociation expressions to a boolean formula of a causal network, we define *expression folding*.

DEFINITION 2.9 (EXPRESSION FOLDING). *Given function $f : \vec{X}' \to \vec{X}$ mapping variables $\vec{X}'$ to $\vec{X}$, the folding $(\mathcal{F}, f)$ of a dissociation expression $\Psi(\vec{X}')$ defines a formula $\Phi = F(\Psi)$, s.t:*

$$\Psi ::= X' \Rightarrow \mathcal{F}(X) = f(X')$$
$$\Psi ::= \sigma(\Psi_1, \Psi_2, \ldots, \Psi_k) \Rightarrow \mathcal{F}(\Psi) = \sigma(\mathcal{F}(\Psi_1), \mathcal{F}(\Psi_2), \ldots, \mathcal{F}(\Psi_k))$$

For example, $f(\{A_1, A_2, B\}) = \{A, A, B\}$ defines a folding $\mathcal{F}$ from $\Psi = A_1 \vee (\bar{A}_2 \wedge B)$ to the formula $\Phi = A \vee (\bar{A} \wedge B)$. In simple terms, a DE $\Psi$ with a folding to $\Phi$, is a representation of $\Phi$ in a larger space of input variables. The use of more inputs captures the distinct effect of variables on the causal path, thus providing the necessary network semantics. We use $|\Psi|$ to denote the cardinality of the input set of $\Psi$. Then $|\Psi| \geq |\Phi|$, and if $|\Psi| = |\Phi|$ then $\Psi = \Phi$.

THEOREM 2.10 (DE MINIMALITY). *If $\mathcal{D}$ the set of all DEs w.r.t. $X_0 \in \vec{X}$ with a folding to $\Phi(\vec{X})$, then $\exists$ unique $\Psi_i \in \mathcal{D}$ of minimum size: $|\Psi_i| = \min_{\Psi \in \mathcal{D}} |\Psi|$ and $\forall j \neq i$, $|\Psi_j| = |\Psi_i| \Rightarrow \Psi_j = \Psi_i$.*

The DE of minimum size replicates those variables, and only those variables, that affect the causal path at more than one location. It is simply called the dissociation expression of $\Phi$, and can be represented as a network (*dissociation network* of $\Phi$), with input nodes $\vec{X}_t$ (Fig. 5b). A folding maps $\vec{X}_t$ back to the original input variables: $\vec{X} = f(\vec{X}_t)$. The reverse mapping is denoted $\vec{X}_t = [\vec{X}]_t = \{X_i \mid f(X_i) \in \vec{X}\}$.

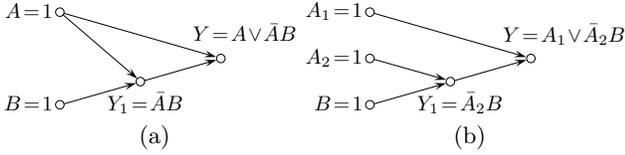

**Figure 5: A causal network $CN$ (a) and its dissociation network $DN$ (b) with respect to $B$.**

DEFINITION 2.11 (FUNCTIONAL CAUSE). *The event $X_i = x_i^0$ is a cause of $\phi$ in a causal model iff:*

FC1. *Both $X_i = x_i^0$ and $\phi$ hold under assignment $\vec{x}^0$*

FC2. *Let $P_\Phi$ and $P_\Psi$ be the potential functions of $\Phi$ and its DE w.r.t. $X_i$, respectively. There exists a support $\vec{S} \subseteq \vec{X} \setminus \{X_i\}$, such that:*
  (a) $\Delta P_\Phi(\vec{S}, X_i) \neq 0$
  (b) $\Delta P_\Psi(\vec{S}'_t) = 0$, *for all subsets $\vec{S}'_t \subseteq [\vec{S}]_t$*

Here, $\Delta P_\Phi(\vec{S}, X_i)$ denotes $\Delta P_\Phi(\vec{S} \cup X_i)$. Condition FC2(b) is analogous to AC2(b) of the HP definition, which requires checking that the effect does not change for all possible combinations of setting the dependent nodes to their original values. Similarly, FC ensures that no part of the changed nodes (the support $\vec{S}$) is counterfactual in the dissociation network. Note that the functional causality definition does not have a minimality condition (equivalent to AC3), as it is directly applied to single literals. As implied by [9] and [12], only primitive events can be causes when dealing with input variables, and therefore a minimality condition is not necessary.

**Intuition.** The definition of functional causes captures three main points: (i) a counterfactual cause is always a cause, (ii) if a variable is not counterfactual under any possible assignment of the other variables, then it cannot be a cause, and (iii) if $X = x^0$ is a counterfactual cause under some assignment that inverts a subset $\vec{S}$ of the other variables, then no part of $\vec{S}$ should be by itself counterfactual.

We revisit the rock thrower example to demonstrate how FC (like AC) can handle preemption. In Sect. 3.2 and Sect. B.1, we show how functional causality successfully handles cases where the HP definition does not give the intuitively correct result, as in Example 2.6 (Shock C).

EXAMPLE 2.12 (TWO THROWERS REVISITED). *The minimal dissociation expression for $\Phi = A \vee (\bar{A} \wedge B)$ with respect to $B$ is $\Psi = A_1 \vee (\bar{A}_2 \wedge B)$, and is depicted in Fig. 5. Then:*

$$P_\Phi = a + b - ab$$
$$P_\Psi = a_1 + b - a_1 b - a_2 b + a_1 a_2 b$$

*For $\vec{S} = \{A\}$, $\Delta P_\Phi(B, \vec{S}) \neq 0$. If $(F, f)$ the folding of $\Psi$ into $\Phi$, then $[\vec{S}]_t = \{A_1, A_2\}$, and $\Delta P_\Psi(A_1) \neq 0$, so $B$ is not a cause.*

Hence, the definition of functional causes effectively capture the difference between the two networks for the two thrower example (Fig. 2) while *only focusing on the input nodes*. In the case of the simple network, $P_\Phi = P_\Psi$ and for $\vec{S} = \{A\}$, $B$ can be shown to be a cause. However, in the more complicated network, the potential function of the dissociation expression gives priority to $A$'s throw and determines that $B$ is not a cause of the bottle breaking.

If the causal network is a tree, then the causal formula is itself a dissociation expression with potential $P_\Phi$. Then, (FC2) simplifies to: (a) $\Delta P_\Phi(\vec{S}, X_i) \neq 0$ and (b) $\forall \vec{S}' \subseteq \vec{S}$ : $\Delta P_\Phi(\vec{S}') = 0$. Causal networks which are trees form an important category of causality problems as they model many practical cases of database queries, and they are characterized by desirable properties, as we show in Sect. 3.4.

**Responsibility.** Responsibility is a measure for degree of causality, first introduced by Chockler and Halpern [6]. We redefine it here for functional causes.

DEFINITION 2.13 (RESPONSIBILITY). *Responsibility $\rho$ of a causal variable $X_i$ is defined as $\frac{1}{|\vec{S}|+1}$ where $\vec{S}$ the minimum support for which $X_i$ is a functional cause of an effect under consideration. $\rho := 0$ if $X_i$ is not a cause.*

Responsibility ranges between 0 and 1. Non-zero responsibility ($\rho > 0$) means that the variable is a functional cause, $\rho = 1$ means it is also a counterfactual cause.

## 3. FORMAL PROPERTIES

Functional causality encodes the *semantics of causal structures* with the help of potential functions which are dependent only on the input variables. In this section we demonstrate that reasoning in terms of *functional causality* provides a more powerful and robust way to reason about causes than *actual causality*. In addition, we give a transitivity result and use it to derive complexity results for certain types of causal network structures.

### 3.1 CC ⊆ FC ⊆ AC

Functional causes are a refined notion of actual causes. Even though the definition of AC does not exclude dependent variables, functional causality does not consider them as possible causes, as their value is fully determined from the input variables. The relationship of functional causality of input variables to actual and counterfactual causality is demonstrated in the following theorem.

THEOREM 3.1 (CC-FC-AC RELATIONSHIP). *Every $X = x^0$ that is a counterfactual cause is also a functional cause, and every $X = x^0$ that is a functional cause is also an actual cause.*

As we have seen with the Shock C example (Example 2.6), the HP definition of *actual causes* is too permissive and determines variables to be causes which should intuitively not be such. The definition of *functional causality* fixes these problems. Appendix B.1 gives a detailed treatment of the Shock C example, both from FC and AC perspectives, and also provides insight into the problems of actual causality.

### 3.2 Causal Network Expansion

Functional, as well as actual causes, rely on the causal network to model a given problem. The two different models of the thrower example displayed in Fig. 2 demonstrate that changes in the network structure can help model priorities of events, which in turn can redefine causality of variables.

In Example 2.5, $B$ is removed as a cause by the addition of an intermediate node in the causal network structure that models the preemption of the effect by node $A$ (Alice's rock is the one that breaks the bottle). This change is also visible in the causal Boolean formula, which is transformed from $\Phi = A \vee B$ to $\Phi_1 = A \vee (\bar{A} \wedge B)$. As we know from Boolean algebra, the two formulas are equivalent as they have the same truth tables. However, they are not *causally equivalent*, as they yield different causality results.

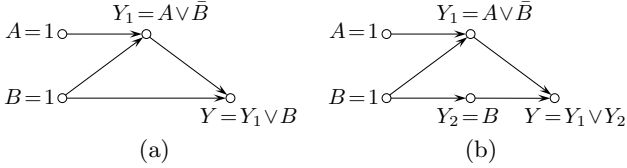

**Figure 6:** Expansion can cause problems for the HP definition: Introducing node $Y_2$ in (b), which merely repeats the value of $B$, does not change function $Y(\vec{X})$, but makes $A$ an actual cause.

Therefore, the grammatical form of the Boolean expression is important in determining causality, and the functional definition captures that through dissociation expressions. It is important to understand how changes in the causal network affect causality, and whether we can state meaningful properties for those changes.

We define *causal network expansion* in a standard way by the addition of nodes and/or edges to the causal structure. A network $CN_e$ with formula $\Phi_e$ is a *node expansion* (respectively *edge expansion*) of $CN$ with formula $\Phi$ if it can be created by the addition of a node (respectively edge) to $CN$, while $\Phi_e \equiv \Phi$. $CN_e$ is a *single-step expansion* if it is either a node or an edge expansion of $CN$.

DEFINITION 3.2 (EXPANSION). *A network $CN_e$ is an expansion of network $CN$ iff $\exists$ set $\{CN_1, CN_2, \ldots, CN_k\}$ with $CN_1 = CN$ and $CN_k = CN_e$, such that $CN_{i+1}$ is a single step expansion of $CN_i$, $\forall i \in [1, k]$.*

Networks represented by the formulas $\Phi_1 = A \vee (\bar{A} \wedge B)$ and $\Phi_2 = (A \wedge \bar{B}) \vee B$ are both expansions of $\Phi = A \vee B$, but note that $\Phi_1$ and $\Phi_2$ are not expansions of one another.

As shown by the thrower example, network expansion can remove causes. As the following theorem states, it can only remove, not add causes.

THEOREM 3.3. *If $CN_e$ with formula $\Phi_e$ is an expansion of $CN$ with formula $\Phi$ and $X_i = x_i^0$ is a cause in $\phi_e$ then $X_i = x_i^0$ is also a cause in $\phi$.*

Specifically in the case where no negation of literals is allowed, changes to the structure do not affect the causality result.

THEOREM 3.4. *If $CN_e$ with formula $\Phi_e$ is an expansion of $CN$ with formula $\Phi$ that does not contain negated variables then $\phi$ and $\phi_e$ have the same causes.*

The properties of formula expansion are important, as they prevent unpredictability due to causal structure changes. Note that the Halpern and Pearl definition does not handle formula expansion as gracefully. Figure 6 demonstrates with an example that the HP definition allows introducing new causes with expansion. $A = 1$ is not a cause in the simple network of Fig. 6a but becomes causal after adding node $Y_2$ in Fig. 6b. Therefore, network expansion is *unpredictable for actual causes*, as there are examples where it can both remove (Fig. 2) or introduce new causes (Fig. 6). This is a strong point for our definition, as causality is tied to the network structure, and erratic behavior due to minor structure changes, as is the case in this example, is troubling.

### 3.3 Functional causes and transitivity

Functional causality only considers *input nodes* in the causal network as permissible causes for events[1]. Under this premise, the notion of *transitivity of causality* is not well-defined, since dependent variables (such as $B$ in the Shock C example 2.6) are never considered permissible causes of events in their descendants. In order to ask the question of transitivity, we allow a dependent variable $Y_1$ to become a possible cause in a *modified causal model* $M'$ with $Y_1$ as additional input variable. We achieve this with the help of an external intervention $[Y_1 \leftarrow y_1^0]$, setting the variable to its actual value $y_1^0$. The new model is then $M' = (\vec{N}, \mathcal{F}')$ with modified structural equations $\mathcal{F}' = \mathcal{F} \setminus \{F_{Y_1}\} \cup \{F'_{Y_1}\}$, where $F'_{Y_1} = y_1^0$, and hence new input variables $\vec{X}' = (\vec{X}, Y_1)$ with original assignment $\vec{x}'^0 = (\vec{x}^0, y_1^0)$.

We can now ask the question of transitivity as follows: Assume that an assignment $X = x^0$ is a cause of $Y_1 = y_1^0$ in a causal model $M$. Further assume that $Y_1 = y_1^0$ is a cause of $Y_2 = y_2^0$ in the modified network $[Y_1 \leftarrow y_1^0]$. Is then $X = x^0$ a cause of $Y_2 = y_2^0$ in the original network $M$? In agreement with recent prevalent (yet not disputed) opinion in causality literature [15, 22], functional causality is not transitive, in general.

COROLLARY 3.5 (NON-TRANSITIVITY). *Functional causality is not transitive, in general.*

Consider again the shock C example 2.6. $A = 1$ is a functional cause of $B = 1$, and $B = 1$ is a functional cause of $C = 1$ in the modified model $[B \leftarrow 1]$. However, $A = 1$ is not a functional cause of $C = 1$ (see Sect. B.1 for details).

Intransitivity of causality is not uncontroversial [19] and humans generally feel a strong intuition that causality *should be* transitive. It turns out that functional causality is actually transitive in an important type of network structure that relates to this intuition: Transitivity holds if there is no causal connection between the original cause ($X$) and the effect ($Y_2$) except through the intermediate node ($Y_1$). This property allows us to deduce a lower complexity for determining causality in restricted settings in Sect. 3.4.

DEFINITION 3.6 (MARKOVIAN). *A node $N$ is Markovian in a causal network $CN$ iff there is no path from any ancestor of $N$ to any descendent of $N$ that does not pass through $N$.*

PROPOSITION 3.7 (MARKOVIAN TRANSITIVITY). *Given a causal model $M$ in which $X = x^0$ is a cause of $Y_1 = y_1^0$ with responsibility $\rho_1$, and in which $Y_1$ is Markovian. Further assume that $Y_1 = y_1^0$ is a cause of $Y_2 = y_2^0$ with responsibility $\rho_2$ in the modified causal model $[Y_1 \leftarrow y_1^0]$. Then $X = x^0$ is a cause of $Y_2 = y_2^0$ in $M$ with responsibility*

$$\rho = (\rho_1^{-1} + \rho_2^{-1} - 1)^{-1}$$

---

[1]This restriction avoids dealing with problematic, inconsistent assignments of variables, which turns out to be one principal reason why the HP definition gives counter-intuitive results. See Appendix B.1 for a detailed discussion.

## 3.4 Complexity

Analogous to Eiter and Lukasiewicz's result that determining actual causes for Boolean variables is NP-hard [9], determining functional causality is also NP-hard, in general.

THEOREM 3.8 (HARDNESS). *Given a Boolean formula $\Phi$ on causal network CN and assignment $\vec{x}^0$ of the input variables, determining whether $X_i = x_i^0$ is a cause of $\phi = \Phi(\vec{x}^0)$ is NP-hard.*

Even though determining functional causality is hard, there are important cases that can be solved in polynomial time.

**Trees.** If the causal network is a tree, then the dissociation network is the same as the causal network and there is a single potential function. Determining causality on a tree can be simplified, as a result of the Markovian transitivity property Prop. 3.7 and the fact that all nodes in a tree are Markovian.

LEMMA 3.9 (CAUSALITY IN TREES). *If $X_i = x_i^0$ is a cause of the output node $Y$ in a tree causal network, and $\vec{p} = \{X, Y_1, Y_2, \ldots, Y\}$ the unique path from $X$ to $Y$, then every node in $\vec{p}$ is a functional cause of all of its descendants in $\vec{p}$. Consequently, $X$ is a cause of all $Y_i \in \vec{p}$.*

Following from Lemma 3.9, causality in cases of tree-shaped causal structures with bounded arity (number of parents per node) is decidable in polynomial time.

THEOREM 3.10 (TREES WITH ARITY $\leq k$). *Given a tree-shaped causal network with formula $\Phi$ and bounded arity and actual assignment $\vec{x}^0$ of the input variables, determining whether $X_i = x_i^0$ is a cause of $\phi = \Phi(\vec{x}^0)$ is in P.*

An even better result is given by Theorem 3.11, that covers the case of causal structures where the function at every node is a primitive boolean operator (AND, OR, NOT), without any restrictions on the arity.

THEOREM 3.11 (TREES WITH PRIMITIVE OPERATORS). *Given a tree causal network with formula $\Phi$ where the function of every node is a primitive boolean operator, i.e. AND, OR, NOT, and assignment $\vec{x}^0$ of the input variables, determining whether $X_i = x_i^0$ is a cause of $\phi = \Phi(\vec{x}^0)$ is in P.*

As demonstrated by Olteanu and Huang in [25], the lineage expressions of safe queries do not have repeated tuples. Lineage expressions for conjunctive queries with no repeated tuples correspond to causal networks that are trees. Following directly from Theorem 3.11, we get complexity results for safe queries.

COROLLARY 3.12 (CAUSES OF SAFE QUERIES). *Determining the causes of safe queries can be done in polynomial time.*

In these tractable cases, due to the transitivity property, responsibility can also be computed in polynomial time, using the formula of Prop. 3.7.

**Positive DNF and CNF.** Another important category of tractable networks are those that correspond to DNF and CNF formulas with no negated literals. This category covers important cases of join queries in a database context.

N(ews feeds)

| nid | story | source |
|---|---|---|
| 1 | ... schools celebrate Indiana's birthday ... | IndyStar |
| 2 | ... economic downturn affected sensitive ... | NYTimes |
| 3 | ... with sequences shot in Indianapolis ... | IndyStar |
| 4 | ... House Approves Bill That Would Ease ... | NYTimes |
| 5 | ... new Bill approved yesterday ... | IndyStar |
| 6 | ... PODS held in Indianapolis this year ... | NYTimes |
| 7 | ... discussed in a recent talk the options to ... | NYTimes |
| 8 | ... Swine Flu Death Toll at 10,000 Since ... | NYTimes |
| 9 | ... Indianapolis welcomes SIGMOD/PODS ... | IndyStar |

F(iltered feed)

| story |
|---|
| ... schools celebrate Indiana's birthday ... |
| ... economic downturn affected sensitive ... |
| ... with sequences shot in Indianapolis ... |
| ... House Approves Bill That Would Ease ... |
| ... PODS held in Indianapolis this year ... |
| ... discussed in a recent talk the options to ... |
| ... Swine Flu Death Toll at 10,000 Since ... |

**Figure 7:** News feed with aggregated data from different sources (above), the filtered feed (below).

THEOREM 3.13 (POSITIVE DNF). *Given a positive DNF formula $\Phi$ and assignment $\vec{x}^0$ of the input variables, determining whether $X_i = x_i^0$ is a cause of $\phi = \Phi(\vec{x}^0)$ is in PTIME.*

THEOREM 3.14 (POSITIVE CNF). *Given a positive CNF formula $\Phi$ and assignment $\vec{x}^0$ of the input variables, determining whether $X_i = x_i^0$ is a cause of $\phi = \Phi(\vec{x}^0)$ is in PTIME.*

## 4. EXPLAINING QUERY RESULTS

In this section, we show how causality can be applied to address examples from the database literature, like provenance and "Why Not?" queries, as well as examples showcasing causality of aggregates. We also demonstrate how our causality framework can model different types of elements that can be considered contributory to a query result, like *query operations* instead of tuples.

### 4.1 WHY SO? and WHY NO?

We revisit our motivating example (Example 1.1), but introduce a slight variation that aggregates data from different news sources to demonstrate how functional causality can be used to answer WHY SO? and WHY NO? questions.

EXAMPLE 4.1 (NEWS AGGREGATOR). *A user has access to the News feed relation N, depicted in Fig. 7. N contains news articles from two different sources, the NY Times and the local IndyStar. The user likes to read the local news from IndyStar, but she prefers the NY Times with regards to broader US or world news. Hence, she does not want to read on topics from IndyStar that are also covered by NY Times. Her filtered feed is constructed by the query*

```
select  N.story
from    N
where   N.source='NYTimes'
        or not exists (
                select *
                from   N as N1
                where  topic(N1.story)=topic(N.story)
                        and N1.source='NYTimes')
```

where `topic()` is a topic extractor modeled as a user-defined function. The user's filtered feed will contain stories from NY Times, and only those stories from IndyStar that NY Times does not cover. Simply, if $S_{NY}$ is an article in NY Times covering a topic, and $S_I$ an article in IndyStar about the same topic, whether the user will see this topic in her feed or not follows a causal model similar to that of Fig. 5a, with boolean formula $\Phi = S_{NY} \vee (\bar{S}_{NY} \wedge S_I)$. The topic appears in F if it appears in either NY Times or IndyStar, but the first gets priority.

When asking what is the cause of getting an article on Indiana's birthday, the user gets tuple 1 from relation N, as it is counterfactual. When asking what is the cause of seeing an article on PODS, she gets the NY Times article (tuple 6), even though IndyStar also had a story about it (tuple 9). The analysis is equivalent to the rock thrower example.

The framework can be used in a similar fashion to respond to "Why No?" questions. Assume tuple $t_{10}$ =(10,'... immigration officials arrest 300...',NYTimes), which was present in yesterday's news feed, but was since then removed. Tuple $t_{10}$ is a functional cause to the WHY NO? question: "Why do I not see news on immigration", as it is counterfactual. Its removal from the feed caused the absence of immigration topics in the user's filtered view.

## 4.2 Aggregates

We next show how functional causality can be applied to determine causes and responsibility for aggregates. We focus here only on positive integers and give complexity results for WHY SO? and WHY NO? for WHY IS SUM $\geq c$? and WHY IS SUM $\not\geq c$?.

**Notation.** Let $\Omega \in \{\text{SUM}, \text{MAX}, \text{AVG}, \text{MIN}, \text{COUNT}\}$ be an aggregate function $\Omega(\vec{V})$ evaluated over a multiset of values $\vec{V}$ from the domain of positive integers, i.e. $v_i \in \mathbb{N}$. Consider a view $R$ with a certain attribute $A$ over which we evaluate the aggregate function. Let $\vec{T}$ be a tuple universe under consideration (i.e. a set of tuples which we consider possible or, simply, the cross product of the active domains for each attribute in $R$), $\vec{T}^+ \subseteq \vec{T}$ the subset of tuples that is in $R$ (i.e. that is true under current assignment) and $\vec{T}^- = \vec{T} - \vec{T}^+$ be those tuples from the tuple universe which are missing (i.e. who are false under current assignment). Denote $\vec{X}$ the vector of Boolean variables where $X_i$ is true or false depending on whether the corresponding tuple $t_i \in T$ is in $T^+$ or not. We write $\Omega(\vec{X})$ as notational shortcut for $\Omega$ evaluated over the subset of $\vec{V}^+ \subseteq \vec{V}$ for which the corresponding Boolean value is true: $\vec{V}^+ = \{v_i \mid v_i \in V \wedge x_i = 1\}$. For example, $\text{SUM}(\vec{x}^0)$ can stand for the query `select SUM(R.A) from R` if $R$ contains tuples with values from $\vec{V}$ in the attribute $R.A$. Let op $\in \{\geq, >, \leq, <, =, \neq\}$. An aggregate condition $\omega^0$ op $c$ for a given constant $c$ is a Boolean expression that is true or false for given assignment $\vec{x}^0$.

DEFINITION 4.2 (WHY SO? AND WHY NO?). Let $\omega^0 = \Omega(\vec{x}^0)$ be the value of an aggregate function for current assignment $\vec{x}^0$. The question of WHY SO? (respectively, WHY NO?) for a condition $\omega^0$ op $c$ that is true (respectively, false) under the current assignment corresponds to the question of which set of tuples $\{t_i\}$ from the tuple universe with original assignment $x_i^0 = 1$ (respectively, 0) is a cause of the event $\phi = (\omega^0 \text{ op } c = \text{true})$ (respectively, false) with responsibility $\rho_i$.

| $R$ | | |
|---|---|---|
| | $A$ | |
| $t_2$ | 20 | $x_2=1$ |
| $t_3$ | 30 | $x_3=1$ |
| $t_5$ | 100 | $x_5=1$ |
| SUM | 150 | |

| $T-R$ | | |
|---|---|---|
| | $A$ | |
| $t_1$ | 10 | $x_1=0$ |
| $t_4$ | 50 | $x_4=0$ |

(a)

| | | WHY SUM$(\vec{x}^0)\geq$ | | | | | WHY SUM$(\vec{x}^0)\not\geq$ | | | |
|---|---|---|---|---|---|---|---|---|---|---|
| $t_i$ | $x_i^0$ | 20 | 30 | 40 | 60 | 130 | 160 | 180 | 210 | 220 |
| $t_1$ | 0 | – | – | – | – | – | 1 | – | $\frac{1}{2}$ | – |
| $t_2$ | 1 | $\frac{1}{3}$ | – | $\frac{1}{2}$ | – | – | – | – | – | – |
| $t_3$ | 1 | $\frac{1}{3}$ | $\frac{1}{2}$ | $\frac{1}{2}$ | – | 1 | – | – | – | – |
| $t_4$ | 0 | – | – | – | – | – | 1 | 1 | $\frac{1}{2}$ | – |
| $t_5$ | 1 | $\frac{1}{3}$ | $\frac{1}{2}$ | $\frac{1}{2}$ | 1 | 1 | – | – | – | – |

(b)

**Figure 8:** Sum example. (a): Relation $R$ with tuples from tuple domain $\vec{T}$. (b): Responsibility $\rho_i$ of $t_i$ for **Why so?** (SUM$(\vec{x}^0)\geq c$) and **Why no?** (SUM$(\vec{x}^0)\not\geq c$).

EXAMPLE 4.3 (SUM EXAMPLE). Consider a tuple universe $\vec{T} = [(10), (20), (30), (50), (100)]$ and a view $R(A)$ with the subset of tuples $\vec{R} = \{(20), (30), (100)\}$. Now consider the query `select SUM(R.A) from R` executed over the view $R$ which returns 150. In our notation, this is represented with a vector $\vec{V} = [10, 20, 30, 50, 100]$, current assignment $\vec{x}^0 = [0, 1, 1, 0, 1]$, and SUM$(\vec{x}^0) = 150$ (see Fig. 8a).

WHY SUM $\geq c$?: $t_3$ is a cause of SUM$(\vec{x}^0) \geq 30$ with responsibility $\frac{1}{2}$. FC2(a): SUM$(\vec{x}^1) \not\geq 30$ for $\vec{x}^1 = [0, 1, \underline{0}, 0, 0]$. FC2(b): SUM$(\vec{x}^{1*}) \geq 30$ for every assignment $\vec{x}^{1*}$ with $x_3^{1*} = 1$ and any subset of $\{x_5^1 = 0\}$ inverted to its original assignment. In contrast, $t_2$ is not a cause: While FC2(a) holds for $\vec{x}^1 = [0, \underline{0}, \underline{0}, \underline{1}, 0]$ with SUM$(\vec{x}^1) \not\geq 30$ (and then $t_2$ would be counterfactual), FC2(b) is not fulfilled for $\vec{x}^{1*} = [0, 1, \underline{0}, 0, 0]$.

WHY SUM $\not\geq c$?: $t_4$ is a cause of $(\text{SUM}(\vec{x}^0) \geq 180) = \text{false}$, as both $x_4$ and the condition are false under current assignment, but would hold for $\vec{x}^1 = [0, 1, 1, \underline{1}, 1]$.

Figure 8b shows responsibility for different values of constant $c$ in Example 4.3 and illustrates that responsibility for SUM is not monotone. In order to compute responsibility for a tuple $t_i$, one must find the smallest set of tuples that, when inverted (i.e. either inserted or deleted) make tuple $t_i$ counterfactual for the condition. We next give complexity results for the SUM aggregator and show that evaluating causality for SUM $\geq c$ is already hard for one relation.

LEMMA 4.4 (SUM POSSIBLE CAUSES). If a tuple $t_i$ is a cause to a WHY SUM $\geq c$? (respectively, WHY SUM $\not\geq c$?) question, then $t_i$ is true (respectively, false) under the actual assignment.

PROPOSITION 4.5 (WHY SO? = WHY NO?). Answers to the question WHY SUM $\geq c$? for an aggregate condition (SUM $\geq c$) = true are the same as WHY SUM $\not\geq c$? for its inverse (SUM $\not\geq c$) = false.

THEOREM 4.6 (SUM HARDNESS). Determining WHY SUM $\geq c$? is NP-complete even for one single input relation.

THEOREM 4.7 (SUM PSEUDO-PTIME). Determining responsibility of a tuple with value $v$ for $(\text{SUM}(\vec{x}^0) \geq c) = \text{true}$ for one single input relation allows a pseudo-polynomial time algorithm $\mathcal{O}\left(n(\omega^0 - c + v)\right)$ where $\omega^0 = \text{SUM}(\vec{x}^0)$.

EXAMPLE 4.8 (NEWS FEED CONTINUED). We will now revisit our motivation example (Example 1.1). The user

| Author | Title | Price | Publisher |
|---|---|---|---|
| | Epic of Gilgamesh | $150 | Hesperus |
| Euripides | Medea | $16 | Free Press |
| Homer | Iliad | $18 | Penguin |
| Homer | Odyssey | $49 | Vintage |
| Hrotsvit | Basilius | $20 | Harper |
| Longfellow | Wreck of the Hesperus | $89 | Penguin |
| Shakespeare | Coriolanus | $70 | Penguin |
| Sophocles | Antigone | $48 | Free Press |
| Virgil | Aeneid | $92 | Vintage |

Figure 9: Books in "Ye Olde Booke Shoppe" [4].

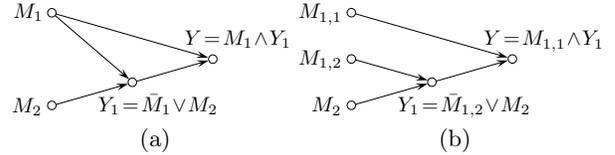

Figure 11: The causal network of Example 4.9 (a), and its DN with respect to $M_2$ (b).

*may be surprised by the increased occurrence of Indianapolis in her personalized feed (5 in total) during a certain week, which is a deviation from the norm. The user can ask a causality query, "Why are there more than 3 occurrences of Indianapolis?". This is a* WHY SO? *query about the* COUNT *on a join between two tables (N and K). The system can calculate the responsibilities of the user's keyword for this aggregate being more than expected. In this case, the responsibilities for the keywords* DB_conf, Purdue *and* Movies *are* $1$, $\frac{1}{2}$ *and* $\frac{1}{2}$ *respectively. This is because* DB_conf *is a counterfactual cause of* COUNT$> 3$, *while the others are causes with support of size 1. This result is intuitive, as there more articles with the* DB_conf *tag (SIGMOD/PODS happening in Indianapolis), than stories with tags* Movies *or* Purdue.

*Similarly, a user may have actually expected to see more news about Indianapolis than the ones she's getting: "Why aren't there more than 6 stories on Indianapolis?". The system can identify the keyword* Indy_500 *as a cause, as it is counterfactual: adding it to the user's keyword list makes the* COUNT *more than 5. Presented with that causality result, the user may decide to include the new keyword in her feed.*

### 4.3 Causes beyond tuples

Provenance and non-answers commonly focus on tuples as discrete units that have contribution to a query result. Our causality framework is not restricted to tuples, but can model any element that could be considered contributory to a result. To showcase this flexibility, we pick an example from Chapman and Jagadish [4] that models *operations* in workflows as possible answers to "Why not?" questions.

EXAMPLE 4.9 (BOOK SHOPPER [4], EX. 1). *A shopper knows that all "window display books" at Ye Olde Booke Shoppe are around $20, and wishes to make a cheap purchase. She issues the query: Show me all window books. Suppose the result from this query is (Euripides, "Medea"). Why is (Hrotsvit, "Basilius") not in the result set? Is it not a book in the book store? Does it cost more than $20? Is there a bug in the query-database interface such that the query was not correctly translated?*

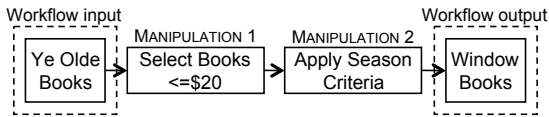

Figure 10: Variation of the query workflow from [4].

Chapman and Jagadish consider a discrete component of a workflow, called *manipulation*, as an explanation of a "Why not?" query. The workflow describing the query of the example is shown in Fig. 10. Roughly, a manipulation is considered *picky* for a non-result if it prunes the tuple. For example, manipulation 1 of Fig. 10 is picky for "Odyssey", as it costs more than $20. Equivalently, a manipulation is *frontier picky* for a set of non-results, if it is the last in the workflow to reject tuples from the set. In this framework, the cause of a non-answer will be a frontier picky manipulation.

In Example 4.9, tuple $t =$(Hrotsvit, "Basilius") passes the price test, but is cut by manipulation 2 as it doesn't satisfy the seasonal criteria. The causal network representing this example is presented in Fig. 11a. Input nodes model the events: $M_1$: manipulation 1 is not potentially picky with respect to $t$, and $M_2$: manipulation 2 is not potentially picky with respect to $t$. At the end, the tuple appears only if neither manipulation is picky: $M_1 \wedge M_2$. Intermediate node $Y_1$ encodes the precedence of the manipulations in the workflow. A tuple will be stopped at point $Y_1$ of the workflow if $M_2$ is picky but $M_1$ was not: $M_1 \wedge \bar{M_2}$. It will pass this point if the opposite holds, so $Y_1 = \overline{M_1 \wedge \bar{M_2}} = \bar{M_1} \vee M_2$, and $Y = M_1 \wedge Y_1$.

Applying the FC framework for $M_1 = 1$ ($M_1$ is not picky), and $M_2 = 0$ ($M_2$ is picky), correctly yields that $M_2$ is the only cause: $\vec{S} = \emptyset$, $\Delta I_\Phi(M_2) \neq 0$. If both manipulations were potentially picky ($M_1 = 0$ and $M_2 = 0$), the FC definition again correctly picks $M_1$ as the only cause with support $\vec{S} = \{M_2\}$ (even though $M_2$ is potentially picky, the tuple never gets to it), which agrees with the WHY NOT? framework that selects as explanation the last manipulation that rejected the tuple.

## 5. RELATED WORK

Our work is mainly related and unifies ideas from three main areas: research on *causality*, *provenance*, and *missing query result explanations*.

**Causality.** Causality is an active research area mainly in logic and philosophy with their own dedicated workshops (see e.g. [1]). The most prevalent definitions of causality are based on the idea of *counterfactual causes*, i.e. causes are explained in terms of counterfactual conditionals of the form *If X had not occurred, Y would not have occurred.* This idea of counterfactual causality can be traced back to Hume [22]. The best known counterfactual analysis of causation in modern times is due to Lewis [18]. In a databases setting, Miklau and Suciu [23] define *critical tuples* as those which can become counterfactual under some value assignment of variables. Halpern and Pearl [13] (HP in short) define a variation they call *actual causality*. Roughly speaking, the idea is that $X$ is a cause of $Y$ if $Y$ counterfactually depends on $X$ under "some" *permissive contingency*, where "some" is elaborately defined. Later, Chockler and Halpern [6] define the degree of *responsibility* as a gradual way to assign causality. Eiter and Lukasiewicz [9] show that the problem of detecting whether $X = x^0$ is an actual cause of an event is $\Sigma_2^P$-complete for general acyclic models and NP-complete for binary acyclic models. They also give an alleged proof showing that actual causality is always reducible to primitive events. However, Halpern [12] later gives an example for non-primitive actual causes, showing this proof to ignore some cases under the original definition. Chockler et al. [7] later apply causality and responsibility to binary Boolean networks, giving a modified definition of cause which, as we show in Sect. B.2, introduces new counter-intuitive problems and, despite claims to be otherwise, is *not equal* to the original HP definition of actual cause.

Our definition of *functional cause* builds upon the HP definition, but extends it with several desirable properties: causes are *always* primitive input variables, network expansion cannot create new causes, and the definition fixes intuitive examples where the HP-definition does not follow consensus in the causality literature. It is these properties that allow us to apply our causality framework to a database setting in Sect. 4.

**Provenance.** Approaches for defining data provenance can be mainly divided into three categories: *how, why*, and *where* provenance [3, 5, 8, 10]. In particular for the "why so" case, we observe a close connection between provenance and causality, where it is often the case that tuples in the provenance for the result of a positive query result are causes. While none of the work on provenance mentions or makes direct connections to causality, those connections can be found. The work by Buneman et al. [3] makes a distinction between why and where provenance that can be connected to causality as follows: *why provenance* returns all tuples that can be considered causes for a particular result, and *where provenance* returns attributes along a particular causal path. Green et al. [10] present a generalization for all types of provenance as semirings; finding functional causes in a Boolean tree, if taken in a provenance context, yields degree-one polynomials for provenance semirings. View data lineage, as presented by Cui et al. [8] also addresses aggregates but lacks a notion of graded contribution and returns all tuples that contribute to an aggregate.

In contrast, our approach can rank tuples according to their *responsibility*, hence our approach allows to determine a gradual contribution with counterfactual tuples ranked first. Also, in contrast to our paper, most of the work on provenance has little or no connection to the philosophical groundwork on causality. We take this work and significantly adapt it so that it can be applied to databases.

**Missing query results.** Very recent work has focused on the question "why no", i.e. why is a certain tuple *not* in the result set? The work by Huang et al. [17] presents provenance for potential answers and never answers. In the case that no insertions or modifications can yield the desired result - usually for privacy or security reasons - the system declares that particular tuple a never answer. Both Huang's work and Artemis [14] handle potential answers by providing tuple insertions or modifications that would yield the missing tuples. Alternatively, Chapman and Jagadish [4] focus on which *manipulation* in the query plan eliminated a specific tuple. Lim et al. [20] adopt a third, explanation-based, approach. This approach aims to answer questions such as *why, why not, how to,* and *what if* for context-aware applications, but does not address a database setting.

Our work, unifies the above approaches in the sense that we model both, *tuples* or *manipulations* as possible causes for missing query answers. Also, our approach unifies the problem of explaining missing query answers (*why* is a tuple not in the query result) with work on provenance (*why* is a tuple in the query result).

**Other.** Minsky and Papert initiated the study of the computational properties of Boolean functions using their representation by polynomials and call this the *arithmetic* instead of the logical form [24, p.27]. This method was later successfully used in complexity theory and became known as *arithmetization* [2].

## 6. CONCLUSIONS AND FUTURE WORK

In this paper, we defined *functional causes*, a rigorous and extensible definition of causality encoding the semantics of *causal structures* with the help of powerful *potential functions*. Through theoretical analysis of its properties, we demonstrated that our definition provides a more powerful and robust way to reason about causes than other established notions of causality. Albeit NP-hard in the general case, common categories of causal networks that correspond to interesting database examples (e.g. safe queries) prove to be tractable. We presented several database examples that portrayed the applicability of our framework in the context of provenance, explanation of non-answers, as well as aggregates. We demonstrated how to determine causes of query results for `SUM` and `COUNT` aggregates, and how these can be ranked according to the causality metric of *responsibility*.

Overall, with this work we establish the theoretical foundations of causality theory in the database context, which we view as a unified framework that deals with query result explanations. It also brings forth many interesting problems that can be explored in future work.

This paper focused on deterministic cases; we plan to extend our framework to *probabilistic data* in the future. The fact that functional causes are based on the use of potential functions makes this extension straightforward: the set of Boolean variables $\vec{X}$ for tuples in the deterministic case becomes a set of probabilities $\vec{x}$. Note that this would not be possible if the causality definition had used just the Boolean formulas. Potential functions also have the additional advantage that they can be analytically manipulated as opposed to Boolean functions. We currently investigate the properties of their derivatives with the intuition that they reveal another facet of causality, particularly with regard to aggregates and probabilities.

**Acknowledgements.** We like to thank Christoph Koch for valuable insights, and Chris Ré for helpful discussions in early stages of this project.

# 7. REFERENCES

[1] International multidisciplinary workshop on causality. IRIT, Toulouse, June 2009. http://www.irit.fr/MICRAC/colloque/articles/extended_abstract_Micrac.pdf.
[2] L. Babai and L. Fortnow. Arithmetization: A new method in structural complexity theory. *Computational Complexity*, 1:41–66, 1991.
[3] P. Buneman, S. Khanna, and W. C. Tan. Why and where: A characterization of data provenance. In *ICDT*, 2001.
[4] A. Chapman and H. V. Jagadish. Why not? In *SIGMOD*, 2009.
[5] J. Cheney, L. Chiticariu, and W. C. Tan. Provenance in databases: Why, how, and where. *Foundations and Trends in Databases*, 1(4):379–474, 2009.
[6] H. Chockler and J. Y. Halpern. Responsibility and blame: A structural-model approach. *J. Artif. Intell. Res. (JAIR)*, 22:93–115, 2004.
[7] H. Chockler, J. Y. Halpern, and O. Kupferman. What causes a system to satisfy a specification? *ACM Trans. Comput. Log.*, 9(3), 2008.
[8] Y. Cui, J. Widom, and J. L. Wiener. Tracing the lineage of view data in a warehousing environment. *ACM Trans. Database Syst.*, 25(2):179–227, 2000.
[9] T. Eiter and T. Lukasiewicz. Complexity results for structure-based causality. *Artif. Intell.*, 142(1):53–89, 2002. Conference version in *IJCAI*, 2002.
[10] T. J. Green, G. Karvounarakis, and V. Tannen. Provenance semirings. In *PODS*, 2007.
[11] N. Hall. Two concepts of causation. In J. Collins, N. Hall, and L. A. Paul, editors, *Causation and Counterfactuals*. MIT Press, 2004.
[12] J. Y. Halpern. Defaults and normality in causal structures. In *KR*, 2008.
[13] J. Y. Halpern and J. Pearl. Causes and explanations: A structural-model approach. Part I: Causes. *Brit. J. Phil. Sci.*, 56:843–887, 2005. Conference version in *UAI*, 2001.
[14] M. Herschel, M. A. Hernández, and W. C. Tan. Artemis: A system for analyzing missing answers. *PVLDB*, 2(2):1550–1553, 2009.
[15] C. Hitchcock. The intransitivity of causation revealed in equations and graphs. *The Journal of Philosophy*, 98(6):273–299, 2001.
[16] M. Hopkins and J. Pearl. Clarifying the usage of structural models for commonsense causal reasoning. In *In Proceedings of the AAAI Spring Symposium on Logical Formalizations of Commonsense Reasoning*, 2003.
[17] J. Huang, T. Chen, A. Doan, and J. F. Naughton. On the provenance of non-answers to queries over extracted data. *PVLDB*, 1(1):736–747, 2008.
[18] D. Lewis. Causation. *The Journal of Philosophy*, 70(17):556–567, 1973.
[19] D. Lewis. Causation as influence. *The Journal of Philosophy*, 97(4):182–197, 2000.
[20] B. Y. Lim, A. K. Dey, and D. Avrahami. *Why* and *why not* explanations improve the intelligibility of context-aware intelligent systems. In *CHI*, 2009.
[21] M. McDermott. Redundant causation. *The British Journal for the Philosophy of Science*, 46(4):523–544, 1995.
[22] P. Menzies. Counterfactual theories of causation. Stanford Encyclopedia of Philosophy, 2008.
[23] G. Miklau and D. Suciu. A formal analysis of information disclosure in data exchange. In *SIGMOD*, 2004.
[24] M. L. Minsky and S. Papert. *Perceptrons - expanded edition: An introduction to computational geometry*. MIT Press, 1987.
[25] D. Olteanu and J. Huang. Secondary-storage confidence computation for conjunctive queries with inequalities. In *SIGMOD*, 2009.
[26] J. Pearl. Comment: Graphical models, causality and intervention. *Statistical Science*, 8(3):266–269, Aug. 1993.
[27] J. Pearl. *Causality: models, reasoning, and inference*. Cambridge University Press, Cambridge, U.K., 2000.

# APPENDIX

## A. NOMENCLATURE

| | |
|---|---|
| $\vec{N}$ | Set of Boolean random variables |
| $\vec{X}$ | Set of input variables |
| $\vec{Y}$ | Set of dependent variables: $\vec{Y} = \vec{N} \setminus \vec{X}$ |
| $M = (\vec{N}, \mathcal{F})$ | Boolean causal Model with nodes $\vec{N} = \vec{X} \cup \vec{Y}$ and functional equations $\mathcal{F} = \{F_N | N \in \vec{N}\}$. $F_X$ for an input variable $X$ is its actual assignment $x^0$. |
| $\vec{x}^0, \vec{y}^0$ | Actual truth assignment of input variables, and resulting truth assignment for dependent variables: $X_i(\vec{x}^0) = x_i^0$, $Y_i(\vec{x}^0) = y_i^0$ |
| $[\vec{S} \leftarrow \vec{s}^1]$ | External intervention replacing the structural equation $F_N$ for each node $N$ in $\vec{S}$ with a truth assignment $n^1$ |
| $CN$ | Causal Network of a causal model |
| $\Phi(\vec{X})$ | Boolean formula for the effect in the $CN$. Corresponds to $Y_j(\vec{X})$ for chosen effect variable $Y_j$ |
| $\phi$ | Effect under consideration. Event of $\Phi(\vec{X}) = \Phi(\vec{x}^0)$, i.e. the effect variable having its actual assignment $\phi = (Y_j = y_j^0)$ |
| $DE, DN$ | Dissociation Expression, Dissociation Network |
| $t: N_1 \to N_2$ | Transformation from network $N_1$ to network $N_2$. $t: CN \to DN$ represents the transformation from a causal to a dissociation network. |
| $[\vec{V}]_t$ | Mapping of a set of nodes $\vec{V}$ from network $N_1$ to $N_2$ under transformation $t: N_1 \to N_2$ |
| $\vec{X}_t, \vec{Y}_t$ | Sets of Boolean variables in $DN$: $\vec{X}_t = [\vec{X}]_t$ and $\vec{Y}_t = [\vec{Y}]_t$ for $t: CN \to DN$ |
| $\vec{x}, \vec{y}$ | Sets of functional variables in $CN$ |
| $\vec{x}_t, \vec{y}_t$ | Sets of functional variables in $DN$ |
| $P_\Phi(\vec{x}), P_\Psi(\vec{x}_t)$ | Potential functions in $CN$ and $DN$, respectively |
| $\Delta P_\Phi(X_i)$ | Change in potential function by inverting input $X_i$: $\Delta P_\Phi(X_i) = P_\Phi(\vec{x}^0) - P_\Phi(1 - x_i^0, \vec{x}^0 \setminus \{x_i\})$ |
| $\vec{x}^0 \oplus \vec{S}$ | the assignment obtained by starting from $\vec{x}^0$ and inverting all variables in $\vec{S}$: $\{1 - s_i^0 \mid S_i \in \vec{S}\} \cup \{\vec{x}^0 \setminus \vec{s}\}$ |
| $\Delta P_\Phi(\vec{S})$ | Change in potential function by inverting all variables in $\vec{S}$: $\Delta P_\Phi(\vec{S}) = P_\Phi(\vec{x}^0) - P_\Phi(\vec{x}^0 \oplus \vec{S})$ |
| $\vec{S}$ | Subset of $\vec{X}$ chosen for condition FC2(b) |
| $[\vec{S}]_t$ | Set $\vec{S}_t \subseteq \vec{X}_t$ that corresponds to $\vec{S} \subseteq \vec{X}$: $\vec{S}_t = \{X_{ij} | X_i \in \vec{S}\}$ |

## B. DETAILS SECTION 2

### B.1 Details on AC and FC for Shock C

In Sect. 2.2, we showed that the HP definition of *actual causes* incorrectly models $A = 1$ to be a cause of $C = 1$ in the Shock C example. We also mentioned but did not show that *functional causes* can model this example correctly. Here, we give the details on this issue. In particular, we show that functional causes can model common sense causality correctly (i.e. $B$'s decision to be mean is a cause for $C$ being shocked) with the help of appropriate *policy variables*, while actual causes cannot, even with the help of more complex network structure[2]. We then give an intuitive explanation of why and where the HP definition of actual causes fails.

Note that the Shock C example [21] is an important example from the philosophical literature that illustrates that causality is not transitive, in general[3]. When philosophers [15, 22] and HP [13] argue for intransitivity of causality, they use examples similar to this one as arguments. Out of the many examples, Shock C is the most compelling case, and the HP definition does not model it correctly. When HP argue for intransitivity of causality, they first have to tweak this example into some modification [13, Example 4.3] where their definition happens to work correctly. In contrast, the definition of functional causes does give the correct attribution of causes given the appropriate network. Also note that the Shock C example is structurally equivalent to the king-assassin-bodyguard example [22, Sec. 4.3], another counterexample to the HP definition.

---
[2]Halpern and Pearl stress the importance of careful causal modeling [13, Sec. 6], implying that actual causes can handle cases correctly given the appropriately modeled network.

EXAMPLE B.1 (SHOCK C - MODEL 1). *In Example 2.6, we used the causal model in Fig. 12, i.e. structural equations*

$$B = A$$
$$C = (A \equiv B) = AB \vee \bar{A}\bar{B}$$

*under actual assignment $A = 1$, and hence $B = 1, C = 1$. In our notation, the set of input variables is $\vec{X} = \{A\}$ and the set of dependent variables $\vec{Y} = \{B, C\}$. The effect $\varphi$ under consideration is $C = 1$.*

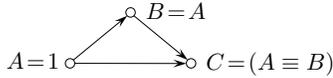

**Figure 12: Model 1: Simple causal model for the Shock C example. Both $A=1$ and $B=1$ are actual causes, neither of them is a functional cause for $C=1$.**

*AC*: *Here, and contrary to common sense, both $A = 1$ and $B = 1$ are actual causes of $C = 1$: (i) $A = 1$ is an actual cause for $\vec{W} = \{B\}$ with $b^1 = 1$. Then AC2(a) holds: $[A \leftarrow 0, B \leftarrow 1] \Rightarrow \neg \phi$. Also AC2(b) holds: $\vec{Z} \setminus \vec{N_c} \setminus \{C\}$ is empty, and $C$ is 1 for either $\vec{W}' = \vec{W}$ or $\vec{W}' = \emptyset$, because of $A \leftarrow 1$. (ii) $B = 1$ is an actual cause for $\vec{W} = \emptyset$.*

*FC*: *There is no functional cause of $C = 1$ as its formula is a tautology. $B$ is a dependent variable and hence $B = 1$ not a permissible cause. $A$ is the only input variable. Since it is not a counterfactual cause and there is no other input variable to invert for $\vec{S}$, it is not a functional cause either.*

The appropriate intuition is that $C = 1$ holds no matter what the assignment of the leaf nodes are. Hence there is no cause. If we want to model $B$'s decision to be mean as a possible cause, we need to model his "intention" with an appropriate *policy variable* as shown next.

---

[3]Several philosophers have taken issue with the idea of causality being intransitive (e.g. [19]) as it seems counter intuitive at first sight. This resonates with Pearl [27, p. 237] asking "why transitivity is so often conceived of as an inherent property of causal dependence". He continues: "One plausible answer is that we normally interpret transitivity to mean the following: If (1) X causes Y and (2) Y causes Z regardless of X, then (3) X causes Z." In Sect. 3.3 we have formalized this observation and given a concrete *Markovian criterium* as sufficient criterium for functional causality to be transitive.

EXAMPLE B.2 (SHOCK C - MODEL 2). *We use the more elaborate causal model from Fig. 13 with structural equations*

$$B = (M \equiv A) = MA \vee \bar{M}\bar{A}$$
$$C = (A \equiv B) = AB \vee \bar{A}\bar{B}$$

*under actual assignment $\vec{x}^0 = \{A = 1, M = 1\}$, and hence $\vec{y}^0 = \{B = 1, C = 1\}$. The intuition is that player $B$ now has the option to be either mean ($M = 1$) and follow the decision of $A$ with $MA$, or not to be mean ($\bar{M}$ or $M = 0$) and do the opposite of $A$, i.e. $\bar{M}\bar{A}$. The motivation is to introduce a new leaf policy variable whose actual assignment $M = 1$ models a permissible modified cause, i.e. $B$'s decision to be mean as a leaf node.*

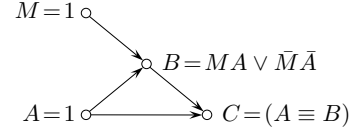

**Figure 13: Model 2: Causal model with explicit *policy variable* $M$ modeling $B$'s decision to be mean. All three $A=1$, $M=1$, and $B=1$ are actual causes for $C$ being shocked according to the HP definition. Only $B$'s decision to be mean ($M=1$) is a functional cause according to our definition, which arguably better represents the common sense interpretation.**

*AC*: *Here again $A = 1$ is an actual cause of $C = 1$. In addition, $M = 1$ and $B = 1$ are causes. Details are the same as in Example B.1.*

*FC*: *$M = 1$ is a functional cause with responsibility 1 (i.e. a counterfactual cause) of $C = 1$, but $A = 1$ is not. (i) $A = 1$ is not a functional cause: First try $\vec{S} = \emptyset$ (equivalent to counterfactual cause). $\Delta P_{\Phi}(A) = 0$, hence it fails FC2(a). Second try $\vec{S} = \{M\}$. Then FC2(a) holds: $\Delta P_{\Phi}(A, M) \neq 0$. However, FC2(b) fails for $S' = S = \{M\}$: $\Delta P_{\Phi}(M) \neq 0$. Hence, $A = 1$ is not a functional cause. (ii) $M = 1$ is a functional cause with responsibility 1, i.e. a counterfactual cause. Inverting $M$ when $\vec{S} = \emptyset$ inverts $C$.*

**Intuition for the AC failure and FC success.** The reasons for the HP definition to give undesired results seem to be twofold: (1) The HP definition allows $\vec{W}$ to be chosen from any node in the causal network, i.e. including nodes in the causal path from the alleged cause $X_i$ to the effect variable $Y_j$; and (2) it allows to give the actual assignment $n_k^1 = n_k^0$ to nodes in $\vec{W}$, i.e. without inverting them. In contrast, our definition of functional causes makes the following changes: (1) we only consider leaf nodes as possible contingencies (i.e. to include in $\vec{S}$). (2) Since we only consider input nodes, we have some implicit minimality criterion for $\vec{S}$. If a variable $X_k$ does not have to be inverted ($x_k^1 = \bar{x}_k^0$) to make another variable a cause, it does not have to be included in $\vec{S}$. (3) We only consider input variables (i.e. leaf nodes) as permissible causes. This has intuitive, practical, and also philosophical appeal: an intermediate dependent variable should not be credited with being cause, it is rather some decision to follow some structural dependency (i.e. some *policy*) rather than another that makes an intermediate node a "visible" cause. As illustrated with Example B.2, we can always introduce new policy variables to a network to analyze the causal effects of structural equations of intermediate nodes[4]. We used this idea in Sect. 4.3 to showcase the why-not approach from Chapman and Jagadish [4] in Example 4.9. Here, we again use this idea to explicitly model $B$'s decision to be mean as an independent input variable in $\vec{X}$, and hence a possible functional cause.

## B.2 The CHK definition for Boolean Circuits

Chockler, Halpern and Kupferman (CHK from now on) give a reformulated definition of actual cause for Boolean circuits with [7, Def. 2.4] and argue that binary acyclic causal models are equivalent to Boolean circuits, i.e. Boolean causal networks where intermediate nodes represent the Boolean operations $\wedge$, $\vee$, or $\neg$, and negations occurs only at the level above the input nodes. As we will show with a simple example, this CHK definition of causality for a Boolean circuit is *not equivalent* to the original HP definition[5].

EXAMPLE B.3 (LOADER [16]). *For a firing squad consisting of shooters $B$ and $C$, it is $A$'s job to load $B$'s gun. In an instance of this problem shown in Fig. 14, $A$ loads $B$'s gun ($A=1$), $B$ does not shoot ($B=0$), but $C$ shoots ($C=1$), and the prisoner dies ($Y=1$).*

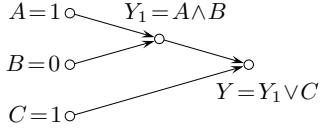

**Figure 14:** $A$ **loads** $B$**'s gun, but** $B$ **does not shoot.** $C$ **shoots and the prisoner** $Y$ **dies.**

<u>AC</u>: *The HP definition (as well as our FC definition) conclude that $A$ is not a cause. This is in accordance with our intuition that $A$ cannot be a cause of the prisoner dying if the gun $A$ loads was not fired.*

<u>CHK</u>: *The causal network of Fig. 14 corresponds to a Boolean circuit with only AND/OR gates. According to [7, Def. 2.4], $A$ is a cause of gate $Y$ if there is an assignment that makes $A$ "critical", i.e. counterfactual in our notation. This assignment exists and is $(b^1=1, c^1=0)$. Hence, according to the CHK definition, $A=1$ is a cause for $Y=1$.*

**Analysis of the CHK definition.** The decisive differences that the CHK definition makes over the original HP definition are twofold: (1) The CHK definition does not inspect the causal path, i.e. the possible changes that a new assignment inflicts to the other gates (e.g. here on $Y_1$). (2) The CHK definition does not check inverting all subsets of $\vec{S}$. For example, in the loader example, the prisoner would not have died for the subset $\vec{S}' = \{c^1 = 0\} \subseteq S$, which indicates that $A=1$ should not be a cause.

While our definition of functional cause also focuses on the input variables only, we made two crucial modifications that avoid new problems such as the loader example, and remedy existing problems of the original HP definition with cases such as the the Shock C example: (1) We use *dissociation expressions* that allow us to manipulate subsets of dissociated input variables while testing causality, and hence, manipulate the relevant causal path only. (2) We test for *all subsets* of the support $\vec{S}$, and hence, verify the causal relevance of the input variable under consideration.

---

[4]This idea of pushing causes to the input nodes seems to be implicit in Pearl [26]. Pearl states that "any external intervention (on a structural function) can be represented graphically as an added parent node".
[5][7, p. 20:6] implies equality to the Boolean formulation of Eiter and Lukasiewicz [9] which is not true.

## B.3 Expression Folding

PROOF THEOREM 2.10 (DE MINIMALITY). For all expressions in $\mathcal{D}$, there exists a folding to $\Phi$. This means that every $\Psi \in \mathcal{D}$ is syntactically equivalent with $\Phi$, but may have one or more instances of variables replaced with new variables. If $\exists \Psi \in \mathcal{D}$ such that $|\Psi| = |\Phi|$, then there is a 1 to 1 correspondence of variables from $\Psi$ to $\Phi$ and therefore $\Psi = \Phi$. Assume $\Psi \in \mathcal{D}$ of minimum size. Obviously, if $|\Psi| = |\Phi|$, then $\forall \Psi' \in \mathcal{D}$ with $|\Psi'| = |\Psi|$, $\Psi' = \Psi$.

We now look at the case where $|\Psi| > |\Phi|$. Assume $\Psi, \Psi' \in \mathcal{D}$ of minimum size, so $|\Psi| = |\Psi'|$. For $\Phi = \sigma(\Phi_1, \Phi_2, \ldots, \Phi_k)$, $\exists \mathcal{F}$ such that $\Phi_i = \mathcal{F}(\Psi_i)$, where $\Psi = \sigma(\Psi_1, \Psi_2, \ldots, \Psi_k)$, and $\exists \mathcal{F}'$ such that $\Phi_i = \mathcal{F}'(\Psi_i')$, where $\Psi' = \sigma(\Psi_1', \Psi_2', \ldots, \Psi_k')$. So $\mathcal{F}(\Psi_i) = \mathcal{F}'(\Psi_i')$. From definition of folding, this holds for all subexpressions $\Phi_i$ of $\Phi$.

If $|\Psi_s'| = |\Psi_s|$ for all subexpressions $\Phi_s$ of $\Phi$, then $\Psi' = \Psi$. Assume $\Phi_s$ some subexpression of $\Phi$ (or $\Phi$ itself), such that $\exists i, j$ such that $|\Psi_{s,i}| < |\Psi_{s,i}'|$ and $|\Psi_{s,j}| > |\Psi_{s,j}'|$, while $|\Psi_s| > |\Psi_s'|$. Such $\Phi_s$ has to exist because $|\Psi'| = |\Psi|$. That means that $\Psi_s \neq \Psi_s'$.

Construct $\Psi_s^* = \sigma(\Psi_{s,1}, \Psi_{s,2}, \ldots, \Psi_{s,j}', \ldots, \Psi_{s,k})$. So $\Psi_s^*$ is the same as $\Psi_s$, apart from subexpression $\Psi_{s,j}$ which is replaced with $\Psi_{s,j}'$. Then $\exists$ folding from $\Psi_s^*$ to $\Phi_s$ and $|\Psi_s^*| < |\Psi_s|$. This means that the DE $\Psi^*$ that results from replacing $\Psi_s$ with $\Psi_s^*$ in $\Psi$ is also a DE for $\Phi$ and $|\Psi^*| < |\Psi|$, which is a contradiction. Therefore $\Psi = \Psi'$. $\square$

## C. DETAILS SECTION 3

### C.1 Functional vs Actual Causes

PROOF THEOREM 3.1 (CC-FC-AC RELASHIONSHIP). If $X_i = x_i^0$ is a counterfactual cause, then it has functional responsibility $\rho = 1$ (for $\vec{S} = \emptyset$, $\Delta P_\Psi(X_i) \neq 0$), and therefore is a FC.

We will show that every FC is an AC. Obviously, condition FC1 implies AC1. We need to show that AC2 holds.

If $X_i = x_i^0$ is a functional cause of effect $\phi$ defined by Boolean formula $\Phi(\vec{X})$, then $\exists \vec{S} \subset \vec{X} \setminus \{X_i\}$ s.t. $\Delta P_\Psi(X_i, \vec{S}) \neq 0$ and $\forall \vec{S}_t' \subseteq [\vec{S}]_t$ $\Delta P_\Psi(\vec{S}_t') = 0$.

We pick $\vec{Z}$ to be the causal path, and $\vec{W}$ the rest of the nodes. Assume $\vec{W}^x = \vec{W} \cap \vec{X}$, $\vec{W}^y = \vec{W} \setminus \vec{W}^x$. We pick assignment $\vec{w}^{x'}$ as follows: $w_j^{x'} = \neg x_j^0$ if $X_j \in \vec{S}$ and $w_j^{x'} = x_j^0$ otherwise. All nodes in $\vec{W}^y$ are descendants of $\vec{W}^x$, so we assign $\vec{w}^{y'}$ as the inferred values from assignment $\vec{w}^{x'}$. $\vec{w}' = \vec{w}^{x'} \cup \vec{w}^{y'}$. From $\Delta P_\Psi(X_i, \vec{S}) \neq 0$ it follows that $\Phi(x_i' \leftarrow \neg x_i^0, \vec{w}') = \neg \phi$, so AC2(a) is satisfied.

Assume some $\vec{W}' \subseteq \vec{W}$ and some $\vec{Z}' \subseteq \vec{Z}$, where $\vec{Z}$ is the causal path of $X_i$ in CN.

Set $\vec{S}_{W'} = \{X_j : X_j \in \vec{W}' \text{ and } w_j' = \neg x_j^0\}$. Obviously, $\vec{S}_{W'} \subseteq \vec{S}$. Also, set $\vec{S}_{Z'} = [\vec{X}]_t \cap ANC(\vec{Z}')$, where $ANC(\vec{Z}')$ the group of all ancestors to any node in $\vec{Z}'$. Finally, set $\vec{S}_t' = \vec{S}_{W'} \setminus \vec{S}_{Z'}$, so $\vec{S}_t' \subseteq [\vec{S}]_t$. Setting $\vec{S}_{Z'}$ to the original values, ensures that all nodes in $\vec{Z}'$ are set to their original values $z^*$. Therefore, $P_\Psi(\neg \vec{s}_{w'}^0, \vec{x}^0 \setminus \vec{s}_{w'}) = \Phi(x_i^0, \vec{W}' \leftarrow \vec{w}', \vec{Z}' \leftarrow \vec{z}^*)$.

Since $\Delta P_\Psi(\vec{S}'_t) = 0$, $\Phi(x_i^0, \vec{W}' \leftarrow \vec{w}', \vec{Z}' \leftarrow \vec{z}^*) = \Phi(\vec{x}^0)$, condition AC2(b) is satisfied.

Condition AC3 is obvious, as $X_i$ is a single literal, and therefore $X_i = x_i^0$ is an actual cause. □

## C.2 Formula Expansion

In this section we give formal definitions of formula expansion.

DEFINITION C.1 (NODE EXPANSION). *Node expansion of a network CN with formula $\Phi$ to a network with formula $\Phi_e$ is the addition of a node $V'$ along an edge $(V, U)$ of the causal network CN, such that $\Phi_e \equiv \Phi$, and none of the formulas of the dependent nodes change.*

DEFINITION C.2 (EDGE EXPANSION). *Edge expansion of a network CN with formula $\Phi$ to a network with formula $\Phi_e$ is the addition of an edge $(V, V')$ in CN, such that $\Phi_e \equiv \Phi$, and none of the formulas of the dependent nodes apart from $V'$ change.*

DEFINITION C.3 (SINGLE STEP EXPANSION). *A network $CN_e$ with formula $\Phi_e$ is a single-step expansion of network CN with formula $\Phi$ if it is either a node or edge expansion of CN.*

DEFINITION C.4 (EXPANSION). *A network $CN_e$ with formula $\Phi_e$ is an expansion of network CN with formula $\Phi$ iff there exists ordered set of networks $\{CN_1, CN_2, \ldots, CN_k\}$ with $CN_1 = CN$ and $CN_k = CN_e$, such that $CN_{i+1}$ is a single step expansion of $CN_i$, $\forall i \in [1, k]$.*

LEMMA C.5. *If $X_i = x_i^0$ is a cause of effect $\phi_e$ in CN with formula $\Phi_e$, which is a single step expansion of formula $\Phi$, then $X_i = x_i^0$ is also a cause of effect $\phi$ in formula $\Phi$.*

PROOF. Assume $\Psi$ the DE of $\Phi$ and $\Psi_e$ the DE of $\Phi_e$. For simplicity we say $X_i$ is a cause, meaning $X_i = x_i^0$ is a cause. $t$ and $T$ represent the dissociation network transformations of $CN$ and $CN^e$ respectively, with respect to $X_i$: $t: CN \to DN$, $T: CN^e \to DN^e$. So, $[\vec{X}]_t$ the input nodes of $DN$, and $[\vec{X}]_T$ the input nodes of $DN^e$. We use the term *potent* to refer to variables in the causal network that map to more than one variable in the dissociation network.

For easiness of representation, we also write $P_\Psi^0(\neg \vec{s}^0)$ to denote $P_\Psi(\neg \vec{s}^0, \vec{x}^0 \backslash \vec{s})$, in other words, all the variables appearing in the argument list of $P_\Psi^0$ are set to the denoted values, and the ones not appearing to their original values given by $\vec{x}^0$.

Since $X_i$ is a cause of $\phi_e$, $\exists \vec{S}$ such that $\Delta P_{\Phi_e}(\vec{S}, X_i) \neq 0$ and $\Delta P_{\Psi_e}(\vec{S}'_T) = 0$, $\forall \vec{S}'_T \subseteq [\vec{S}]_T$.

By definition of expansion, $\Phi_e(\vec{X}) = \Phi(\vec{X})$, and therefore $P_\Phi = P_{\Phi_e}$, which means that $\Delta P_\Phi(\vec{S}, X_i) \neq 0$ for the same set $\vec{S}$.

If $\Phi_e$ is a node expansion of $\Phi$, then $P_\Psi = P_{\Psi_e}$, and therefore $X_i$ is a cause of $\phi$.

If $\Phi_e$ is an edge expansion of $\Phi$, by the addition of an edge $(v, u)$, then node $v$ may become potent with respect to $X_i$ in $\Phi_e$. If $v$ is not potent, then $[X]_t = [X]_T$, and therefore $P_{\Psi_e} = P_\Psi$, which means that $X_i$ is also a cause of $\phi$.

If $v$ is potent with respect to $X_i$, then $DN^e$ contains a set of replicated nodes $\vec{V}'$ of $\vec{V}$ (node $v$ and its ancestors), which are not contained in $DN$, so $[X]_t \subset [X]_T$. Denote as $\vec{X}_v$ the subset of $\vec{X}$ that are ancestors of $v$, and $\vec{X}_{v'}$ the subset of $\vec{X}$ that are ancestors of the replica $v'$ in $DN^e$. Then $\vec{X}_T = \vec{X}_t \cup \vec{X}_{v'}$.

If $\vec{X}_v \cap [\vec{S}]_T = \emptyset$, then $\{\vec{X}_v \cup \vec{X}_{v'}\} \cap \vec{S}_T = \emptyset$. Then, $\vec{S}_t = \vec{S}_T$, and for any $\vec{S}'_t \subseteq \vec{S}_t$, $\vec{S}'_t$ is also a subset of $\vec{S}_T$. That means that $P_\Psi^0(\neg \vec{s}'^0_t) = P_{\Psi_e}^0(\neg \vec{s}'^0_t) = P_{\Psi_e}(\vec{x}^0)$, which means that $\Delta P_\Psi(\vec{s}'_t) = 0$.

If $\vec{X}_v \cap \vec{S}_T \neq \emptyset$, then $\vec{X}_{v'} \cap \vec{S}_T \neq \emptyset$. Then $\vec{S}_t \subset \vec{S}_T$, as $\vec{X}_v \cap \vec{S}_t \neq \emptyset$, but $\vec{X}_{v'} \cap \vec{S}_t = \emptyset$. For any $\vec{S}'_t \subseteq \vec{S}_t$, $\vec{X}'_v = \vec{S}'_t \cap \vec{X}_v$, and $\vec{X}'_{v'}$ its replicated equivalent in $DN^e$. $\exists \vec{S}'_T = \vec{S}'_t \cup \vec{X}'_{v'}$, and $\vec{S}'_T \subseteq \vec{S}_T$. Also, by definition of expansion,

$$P_\Psi^0(\neg \vec{x}'^0_v, \neg \vec{s}'^0_t \backslash \vec{x}'_v) =$$
$$= P_{\Psi_e}^0(\neg \vec{x}'^0_v, \neg \vec{x}'^0_{v'}, \neg \vec{s}'^0_t \backslash \vec{x}'_v) = P_{\Psi_e}(\vec{x}^0)$$

Therefore, $\Delta P_\Psi(\vec{S}'_t) = 0$, $\forall \vec{S}'_t \subset \vec{S}_t$ in all cases of expansion, which means that $X_i$ is a cause of $\phi$. □

LEMMA C.6. *If $X_i = x_i^0$ is a cause of effect $\phi$ in CN with formula $\Phi$, and $\Phi_e$ a single step expansion of $\Phi$ that does not contain negated variables, then $X_i = x_i^0$ is also a cause of effect $\phi_e$ formula $\Phi_e$.*

PROOF. For the most part, this proof is similar to the proof of Lemma C.5.

Assume $\Psi$ the DE of $\Phi$ and $\Psi_e$ the DE of $\Phi_e$. For simplicity we say $X_i$ is a cause, meaning $X_i = x_i^0$ is a cause. $t$ and $T$ represent the dissociation network transformations of $CN$ and $CN^e$ respectively, with respect to $X_i$: $t: CN \to DN$, $T: CN^e \to DN^e$. So, $[\vec{X}]_t$ the input nodes of $DN$, and $[\vec{X}]_T$ the input nodes of $DN^e$. We use the term *potent* to refer to variables in the causal network that map to more than one variable in the dissociation network.

For easiness of representation, we also write $P_\Psi^0(\neg \vec{s}^0)$ to denote $P_\Psi(\neg \vec{s}^0, \vec{x}^0 \backslash \vec{s})$, in other words, all the variables appearing in the argument list of $P_\Psi^0$ are set to the denoted values, and the ones not appearing to their original values given by $\vec{x}^0$.

Since $X_i$ is a cause of $\phi$, $\exists \vec{S}$ such that $\Delta P_\Phi(\vec{S}, x_i) \neq 0$ and $\Delta P_\Psi(\vec{S}'_t) = 0$, $\forall \vec{S}'_t \subseteq [\vec{S}]_t$.

By definition of expansion, $\Phi_e(\vec{X}) = \Phi(\vec{X})$, and therefore $P_\Phi = P_{\Phi_e}$, which means that $\Delta P_{\Phi_e}(\vec{S}, X_i) \neq 0$ for the same set $\vec{S}$.

If $\Phi_e$ is a node expansion of $\Phi$, then $P_\Phi = P_{\Phi_e}$ and $P_\Psi = P_{\Psi_e}$, and therefore $X_i$ is a cause of $\phi_e$.

If $\Phi_e$ is an edge expansion of $\Phi$, by the addition of an edge $(v, u)$ node $v$ may become potent with respect to $X_i$ in $\Phi_e$. If $v$ is not potent, then $[X]_t = [X]_T$, and therefore $P_{\Psi_e} = P_\Psi$, which means that $X_i$ is also a cause of $\phi_e$.

If $v$ is potent with respect to $X_i$, then $DN^e$ contains a set of replicated nodes $\vec{V}'$ of $\vec{V}$ (node $v$ and its ancestors), which are not contained in $DN$, so $[X]_t \subset [X]_T$. Denote as $\vec{X}_v$ the subset of $\vec{X}$ that are ancestors of $v$, and $\vec{X}_{v'}$ the subset of $\vec{X}$ that are ancestors of the replica $v'$ in $DN^e$. Then $\vec{X}_T = \vec{X}_t \cup \vec{X}_{v'}$.

If $\vec{X}_v \cap [\vec{S}]_t = \emptyset$, then $\{\vec{X}_v \cup \vec{X}_{v'}\} \cap [\vec{S}]_T = \emptyset$. Then, $\vec{S}_T = \vec{S}_t$, and for any $\vec{S}'_t \subseteq \vec{S}_t$, $\vec{S}'_t$ is also a subset of $\vec{S}_t$. That means that $P_{\Psi_e}^0(\neg \vec{s}'^0_t) = P_\Psi^0(\neg \vec{s}'^0_t) = \Phi_e(\vec{x}^0)$, which means that $\Delta P_{\Psi_e}(\vec{S}'_t) = 0$.

If $\vec{X}_v \cap \vec{S}_t \neq \emptyset$, then $\vec{X}_v \cap \vec{S}_T \neq \emptyset$ and $\vec{X}_{v'} \cap \vec{S}_T \neq \emptyset$. For any $\vec{S}'_T \subseteq \vec{S}_T$, denote $\vec{X}'_v = \vec{X}_v \cap \vec{S}'_T$ and $\vec{X}'_{v'} = \vec{X}_{v'} \cap \vec{S}'_T$. Assume $\vec{X}_c \subseteq \vec{X}'_v$ is the set of all variables in $\vec{X}'_v$ that have an equivalent in $X'_{v'}$, i.e. their replicas are in $\vec{X}'_{v'}$. Then $\vec{X}'_v$ and $\vec{X}'_{v'}$ can be rewritten as follows: $\vec{X}'_v = \vec{X}_1 \cup \vec{X}_c$ and $\vec{X}'_{v'} = \vec{X}'_2 \cup \vec{X}'_c$. The replicas of $\vec{X}_1$, $\vec{X}_2$, and $\vec{X}_c$ in $DN^e$ are $\vec{X}'_1$, $\vec{X}'_2$ and $\vec{X}'_c$ respectively, so $\vec{X}'_1 \not\subset \vec{X}'_{v'}$ and $\vec{X}_2 \not\subset \vec{X}'_v$.

Also, denote $\vec{S}''_T = \vec{S}'_T \backslash \{\vec{X}'_v \cup \vec{X}'_{v'}\}$.

$$\Delta P_{\Psi_e}(\vec{S}'_t) = P_{\Psi_e}(\vec{x}^0) - P^0_{\Psi_e}(\neg \vec{x}^0_1, \neg \vec{x}^0_c, \neg \vec{x}'^0_2, \neg \vec{x}'^0_c, \neg \vec{s}''^0_T)$$

Also, we know that $\Delta P_{\Psi}(\vec{s}'_t) = 0$, $\forall \vec{S}'_t \subseteq \vec{S}_t$. Assign $S'_t = \vec{X}_1 \cup \vec{X}_2 \cup \vec{X}_c \cup \vec{S}''_t$ and $\vec{S}''_t = \vec{S}'_t \backslash \{\vec{X}_1 \cup \vec{X}_2 \cup \vec{X}_c\}$. Then $S'_t \subseteq \vec{S}_t$, and therefore:

$$P^0_\Psi(\neg \vec{x}^0_1, \neg \vec{x}^0_2, \neg \vec{x}^0_c, \neg \vec{s}''^0_t) = P_\Psi(\vec{x}^0)$$

By definition of expansion:

$$P^0_{\Psi_e}(\neg \vec{x}^0_1, \neg \vec{x}^0_2, \neg \vec{x}^0_c, \neg \vec{x}'^0_1, \neg \vec{x}'^0_2, \neg \vec{x}'^0_c, \neg \vec{s}''^0_t) = $$
$$= P^0_\Psi(\neg \vec{x}^0_1, \neg \vec{x}^0_2, \neg \vec{x}^0_c, \neg \vec{s}''^0_t) \Rightarrow$$

$$P^0_{\Psi_e}(\neg \vec{x}^0_1, \neg \vec{x}^0_2, \neg \vec{x}^0_c, \neg \vec{x}'^0_1, \neg \vec{x}'^0_2, \neg \vec{x}'^0_c, \neg \vec{s}''^0_t) = P_\Psi(\vec{x}^0)$$

We compute $P^0_{\Psi_e}(\neg \vec{x}^0_1, \neg \vec{x}^0_c, \neg \vec{x}'^0_2, \neg \vec{x}'^0_c, \neg \vec{s}'^0_T)$, re-written as: $P^0_{\Psi_e}(\neg \vec{x}^0_1, \vec{x}^0_2, \neg \vec{x}^0_c, \vec{x}'^0_1, \neg \vec{x}'^0_2, \neg \vec{x}'^0_c, \neg \vec{s}''^0_t)$.

Since variables are not negated, $P_\Psi$ and $P_{\Psi_e}$ are monotonous. Therefore:

$$P_\Psi(\vec{x}^0) = P^0_{\Psi_e}(\neg \vec{x}^0_1, \neg \vec{x}^0_2, \neg \vec{x}^0_c, \neg \vec{x}'^0_1, \neg \vec{x}'^0_2, \neg \vec{x}'^0_c, \neg \vec{s}''^0_t) \le$$
$$\le P^0_{\Psi_e}(\neg \vec{x}^0_1, \vec{x}^0_2, \neg \vec{x}^0_c, \vec{x}'^0_1, \neg \vec{x}'^0_2, \neg \vec{x}'^0_c, \neg \vec{s}''^0_t) \le$$
$$\le P^0_{\Psi_e}(\vec{x}^0_1, \vec{x}^0_2, \vec{x}^0_c, \vec{x}'^0_1, \vec{x}'^0_2, \vec{x}'^0_c, \neg \vec{s}''^0_t) =$$
$$P^0_\Psi(\vec{x}^0_1, \vec{x}^0_2, \vec{x}^0_c, [\neg \vec{s}''^0_t]_{\succ}) = P_\Psi(\vec{x}^0)$$

Therefore, $P^0_{\Psi_e}(\neg \vec{x}^0_1, \neg \vec{x}^0_c, \neg \vec{x}'^0_2, \neg \vec{x}'^0_c, \neg \vec{s}''^0_T) = P_\Psi(\vec{x}^0)$, which means that $\Delta P_{\Psi_e}(\vec{s}'_t) = 0$, and $X_i$ is a cause of $\phi_e$. $\square$

These two lemmas lead to the general theorems of formula expansion presented in Sect. 3.2.

PROOF THEOREM 3.3 (FORMULA EXPANSION). Since $\Phi_e$ is an expansion of $\Phi$, $\exists$ ordered set of formulas $\{\Phi_1, \Phi_2, \ldots, \Phi_k\}$ with $\Phi_1 = \Phi$ and $\Phi_k = \Phi_e$, such that $\Phi_{i+1}$ is a single step expansion of $\Phi_i$, $\forall i \in [1, k]$.

As shown in Lemma C.5, if $X_i = x^0_i$ is a cause of $\phi^i$ then it is also a cause of $\phi_{i-1}$, $\forall i \in [2, k]$. Therefore, if $X_i = x^0_i$ a cause of $\phi_k$, it is also a cause of $\phi_1$. $\square$

PROOF THEOREM 3.4 (EXP. OF POSITIVE FORMULAS). As shown by Theorem 3.3, if something is a cause of $\phi_e$, then it is also a cause of $\phi$. We now need to show that if $X_i = x^0_i$ is a cause of $\phi$, then it is also a cause of $\phi_e$.

Since $\Phi_e$ is an expansion of $\Phi$, $\exists$ ordered set of formulas $\{\Phi_1, \Phi_2, \ldots, \Phi_k\}$ with $\Phi_1 = \Phi$ and $\Phi_k = \Phi_e$, such that $\Phi_{i+1}$ is a single step expansion of $\Phi_i$, $\forall i \in [1, k]$.

As shown in Lemma C.6, if $X_i = x^0_i$ is a cause of $\Phi^{i-1}$ then it is also a cause of $\Phi_i$, $\forall i \in [2, k]$. Therefore, if $X_i = x^0_i$ a cause of $\Phi_1$, it is also a cause of $\Phi_k$. $\square$

## C.3 Markovian transitivity

PROOF PROP. 3.7 (MARKOVIAN TRANSITIVITY). Here we denote as $I_{Y_i}$ the potential function of formula $\Phi_{Y_i}$, and $P_{Y_i}$ the potential function of the DE of $\Phi_{Y_i}$. To simplify notation, we omit the non-negated terms in the potential functions. So we write $P(\neg \vec{s}^0)$ meaning $P(\neg \vec{s}^0, \vec{x}^0 \backslash \vec{s}^0)$.

Assume $X$ is a functional cause of $Y_1$ with responsibility $\rho_1$. Then there exists $\vec{S}_{Y_1} \subseteq \vec{X} \cap ANC(Y_1)$, such that $\Delta I_{Y_1}(X, \vec{S}_{Y_1}) \ne 0$, and $\forall \vec{S}'_t \subseteq [\vec{S}_{Y_1}]_t$, $\Delta P_{Y_1}(\vec{S}'_t) = 0$, and $\rho_1 = \frac{1}{|\vec{S}_{Y_1}|+1}$.

Also, in the mutilated network, $Y_1$ is a cause of a $Y_2$ with responsibility $\rho_2$, then $\exists$ a minimum set $\vec{S}_{Y_2} \subseteq \vec{X} \cap \{ANC(Y_2) \backslash ANC(Y_1)\}$, such that $\Delta I_{Y_2}(X, \vec{S}_{Y_2}) \ne 0$, and $\forall \vec{S}'_t \subseteq [\vec{S}_{Y_2}]_t$ $\Delta P_{Y_2}(\vec{S}'_t) = 0$, where $I_{Y_2}$ and $P_{Y_2}$ the potential functions of $Y_2$ in the mutilated CN and corresponding DN. Also, $\rho_2 = \frac{1}{|\vec{S}_{Y_2}|+1}$. Obviously, $\vec{S}_{Y_1} \cap \vec{S}_{Y_2} = \emptyset$ and $[\vec{S}_{Y_1}]_t \cap [\vec{S}_{Y_2}]_t = \emptyset$. Since $Y_1$ is markovian, no ancestors of $Y_1$ connect to the rest of the network without going through $Y_1$. Therefore, nodes in $[\vec{S}_{Y_1}]_t$ also do not connect to the rest of the network without going through $Y_1$.

Assume $\Phi$ is the Boolean formula at $Y_2$ on the complete network. Also, denote $\vec{X}_{Y_1} = \{\vec{X} \backslash X\} \cap ANC(Y_1)$ and $\vec{X}_{Y_2} = \{\{\vec{X} \backslash X\} \cap ANC(Y_2)\} \backslash \vec{X}_{Y_1}$. Then $I_\Phi(\vec{x}) = I_\Phi(x, \vec{x}_{Y_1}, \vec{x}_{Y_2}) = I_{Y_2}(I_{Y_1}(x, \vec{x}_{Y_1}), \vec{X}_{x_2})$. Similarly, in the DN we get: $P_\Phi(\vec{x}) = P_{Y_2}(P_{Y_1}(x, \vec{x}_{Y_1}), \vec{x}_{Y_2})$.

Set $\vec{S} = \vec{S}_{Y_1} \cap \vec{S}_{Y_2}$. $I_\Phi(\neg x^0, \neg \vec{s}^0) = I_\Phi(\neg x^0, \neg \vec{s}^0_{Y_1}, \neg \vec{s}^0_{Y_2}) = I_{Y_2}(I_{Y_1}(\neg x^0, \neg \vec{s}^0_{Y_1}), \neg \vec{s}^0_{Y_2}) = I_{Y_2}(\neg y^0_1, \neg \vec{s}^0_{Y_2}) = \neg y^0_2$. Therefore, $\Delta I_\Phi(X, \vec{S}) \ne 0$.

Assume set $\vec{S}'_t \subseteq [\vec{S}]_t$. Set $\vec{S}_1 = \vec{S}'_t \cap [\vec{S}_{Y_1}]_t$ and $\vec{S}_2 = \vec{S}'_t \cap [\vec{S}_{Y_2}]_t$. Clearly, $\vec{S}_1 \cap \vec{S}_2 = \emptyset$. Then, $P_\Phi(\neg \vec{s}'^0_t) = P_\Phi(\neg \vec{s}^0_1, \neg \vec{s}^0_2) = P_{Y_2}(P_{Y_1}(\neg \vec{s}^0_1), \neg \vec{s}^0_2) = P_{Y_2}(y^0_1, \neg \vec{s}^0_2) = y^0_2$. Therefore, $\Delta P_\Phi(\vec{S}'_t) = 0$ for any $\vec{S}'_t$. Therefore, $X$ is a functional cause of $Y_2$.

$\vec{S}$ is also minimal, as $\vec{S}_{Y_1}$ and $\vec{S}_{Y_2}$ are minimal and disjoint. Therefore, $X$ is a cause of $Y_2$ with responsibility

$$\rho = \frac{1}{|\vec{S}|+1} = \frac{1}{|\vec{S}_{Y_1}| + |\vec{S}_{Y_2}|+1} = (\rho_1^{-1} + \rho_2^{-1} - 1) \quad \square$$

## C.4 Complexity of functional cause

PROOF THEOREM 3.8 (COMPLEXITY). In this proof, we denote with $I_\Phi$ the potential function of a formula $\Phi$, and with $P_\Phi$ the potential function of the DE of $\Phi$. Note that $\Psi$ represents a 3DNF formula, and not a DE of $\Phi$.

We use a reduction, inspired by the proof [9, Theorem 3.3], from the *non-tautology problem of a 3DNF*: given a 3DNF propositional formula $\Psi$ over a set of variables $\vec{X} = \{X_1, \ldots, X_n\}$, is there a truth assignment for $\vec{X}$ that makes $\Psi$ false.

We transform an instance of the 3DNF tautology problem to a problem of determining whether a variable is a functional cause as follows. We create a dependent variable for every conjunct $C_j$ in $\Psi$ with $\vec{C} = \{C_1, \ldots, C_k\}$. Every variable $X_i$ connects to every $C_j$ it is part of. Eventually every $C_j$ has 3 incoming edges. We also create a separate input node $X_0$ and an output node $Y$ with incoming edges from $X_0$ and all the $C_j \in \vec{C}$, which applies the OR function to its inputs. The final output node $Y$ has formula $Y = X_0 \vee C_1 \vee \ldots \vee C_k = X_0 \vee \Psi$ (see Fig. 15).

Assume initial assignment $X_0 = 1$ and any assignment $\vec{x}^0$ for $\vec{X}$. Also name $\Phi$ the Boolean formula representing node $Y$. We will show that $\Psi$ is a tautology, iff $X_0 = 1$ is not a cause of $Y = 1$. (1) If $\Psi$ is a tautology, then $Y = 1$ for all assignments, and therefore $\not\exists \vec{S}$ such that $\Delta I_\Phi(X_0, \vec{S}) \ne 0$, as the potential function of the formula is also always 1. (2)

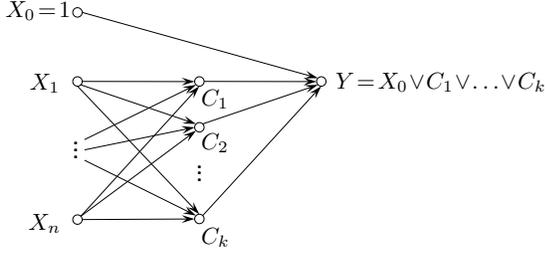

**Figure 15:** Reduction from 3DNF tautology to determining functional causes in a Boolean network.

If $\Psi$ is not a tautology, then there exists an assignment $\vec{x}^1$ for which $\Psi(\vec{x}^1) = 0$. We assign $\vec{S} = \{X_i \,|\, x_i^1 \neq x_i^0\}$. Then, clearly, $\Delta I_\Phi(X_0, \vec{S}) \neq 0$, as $I_\Phi(\neg x_0, \neg \vec{s}^0, \vec{x}^0 \backslash \vec{s}) = 0$. Also, for any $S'$, $\Delta P_\Phi(\vec{S}) = 0$, as $\Phi = 1$ for $X_0 = 1$. Therefore, if $\Psi$ is not a tautology, $X_0 = 1$ is a cause.

By this, we have shown that determining whether a variable is a functional cause of an event in a Boolean causal network is NP-hard in the size of the network. □

PROOF LEMMA 3.9 (CAUSALITY IN TREES). If the causal network is a tree, then every node is Markovian. That means that there is a single path from any variable $X_i$ to the effect variable $Y$, consisting of variables $\vec{p} = \{Y_1, \ldots, Y\}$.

In a tree causality is transitive (Prop. 3.7): if $X_i$ is a cause of $Y_j \in \vec{p}$, and $Y_j$ is a cause of $Y$, then $X_i$ is a cause of $Y$.

We will show that if $X_i$ is not a cause of $Y_j \in \vec{p}$, then $X_i$ cannot be a cause of $Y$.

Assume that $X_i$ is a cause of $Y$. Then $\exists \vec{S}$ such that $\Delta P_Y(X_i, \vec{S}) \neq 0$ and $\forall \vec{S}' \subseteq \vec{S}$, $\Delta P_Y(\vec{S}') = 0$. Set $\vec{S}_Y = \vec{S} \cap A(Y_j)$, where $A(Y_j)$ the subset of $\vec{X}$ that are ancestors of $Y_j$. Also set $\vec{S}_R = \vec{S} \backslash \vec{S}_Y$.

If $\vec{S}$ is inverted, $X_i$ is counterfactual for $Y$:
$$P_{Y_k}(\neg x_i^0, \neg \vec{s}^0, \vec{x}^0 \backslash \{x_i, \vec{s}\}) \neq P_Y(x_i^0, \neg \vec{s}^0, \vec{x}^0 \backslash \{x_i, \vec{s}\})$$

Since there is only one path from $X_i$ to $Y$ through $Y_j$, then $Y_j$ has to also flip values when $\vec{S} \leftarrow \neg \vec{s}^0$ and $X_i \leftarrow \neg x_i^0$, meaning $\Delta P_{Y_j}(X_i, \vec{S}_Y) = \Delta P_{Y_j}(X_i, \vec{S}) \neq 0$. Also, $Y_j$ should have its original value when $X_i$ is set to its original value with $\vec{S}$ inverted, otherwise $X_i$ would not be counterfactual for $Y$. So, $\Delta P_{Y_j}(\vec{S}_Y) = 0$.

Assume $\vec{S}_Y' \subseteq \vec{S}_Y$. $\Delta P_Y(\vec{S}_Y' \cap \vec{S}_R) = 0$, which means that $\Delta P_{Y_j}(\vec{S}_Y') = 0$.

Therefore, $Y_j$ also has to be a cause of $Y$. □

PROOF THEOREM 3.10 (RESTRICTED ARITY). Follows directly from Lemma 3.9. Since the tree has restricted arity $\leq k$, determining causality of a node for its immediate descendant is polynomial. Also, because of transitivity, as shown in Lemma 3.9 to show that $X$ is a cause of $Y$ it suffices to show that every node in the path $\vec{p} = \{Y_1, \ldots, Y\}$ is a cause to its immediate descendant. The length of the path grows with $\log n$, and therefore determining whether $X$ is a cause of $Y$ is in P. □

PROOF THEOREM 3.11 (PRIMITIVE OPERATORS). Assume $Y_j$ is the immediate descendant of $Y_i$ in a tree causal network. If the function of $Y_j$ is a primitive boolean operator, i.e. AND, OR, NOT, it can be decided in polynomial time if $Y_i$ is a cause of $Y_j$.

**Case A:** $Y_j$ is an AND node

Set $\vec{S}_Y = parents(Y_j) \backslash \{Y_i\}$. If $y_j^0 = 1$, then $Y_i$ is a cause because it is counterfactual. If $y_j^0 = 0$ and $y_i^0 = 1$, then $Y_i$ is not a cause because the AND function is monotone, so setting $Y_i$ to 0 will not invert $Y_j$ under any contingency. If $y_j^0 = 0$ and $y_i^0 = 0$, $\exists \vec{S} \subseteq \vec{X}$ that sets $\vec{S}_Y$ to true. This is always possible because in a tree every input node participates in the formula exactly once. Therefore, $Y_i$ becomes counterfactual for $\vec{S}$, and $Y_j$ is always 0 when $Y_i$ is set to $y_i^0 = 0$, which means that $\Delta P_\Phi(\vec{S}_t') = 0$ for any $S_t'$. This means that when $y_j^0 = 0$ and $y_i^0 = 0$, $Y_i$ is a cause of $Y_j$. Therefore, it is determined in constant time whether a node is a cause of its immediate descendant AND node.

**Case B:** $Y_j$ is an OR node

Set $\vec{S}_Y = parents(Y_j) \backslash \{Y_i\}$. If $y_j^0 = 0$, then $Y_i$ is a cause because it is counterfactual. If $y_j^0 = 1$ and $y_i^0 = 0$, then $Y_i$ is not a cause because the OR function is monotone, so setting $Y_i$ to 1 will not invert $Y_j$ under any contingency. If $y_j^0 = 1$ and $y_i^0 = 1$, $\exists \vec{S} \subseteq \vec{X}$ that sets $\vec{S}_Y$ to false. This is always possible because in a tree every input node participates in the formula exactly once. Therefore, $Y_i$ becomes counterfactual for $\vec{S}$, and $Y_j$ is always 1 when $Y_i$ is set to $y_i^0 = 1$, which means that $\Delta P_\Phi(\vec{S}_t') = 0$ for any $S_t'$. This means that when $y_j^0 = 1$ and $y_i^0 = 1$, $Y_i$ is a cause of $Y_j$. Therefore, it is determined in constant time whether a node is a cause of its immediate descendant OR node.

**Case B:** $Y_j$ is an NOT node
Causality can be determined in constant time because there is a single input to the node.

Therefore it is decidable in constant time whether $Y_i$ is a cause of its immediate descendant $Y_j$. That means that to show that $X$ is a cause of $Y$ it suffices to show that every node in the path $\vec{p} = \{Y_1, \ldots, Y\}$ is a cause to its immediate descendant, each of which steps can be done in constant time. The length of the path grows with $\log n$, and therefore determining whether $X$ is a cause of $Y$ is in P. □

PROOF THEOREM 3.13 (POSITIVE DNF). Assume $\Phi$ a formula in DNF with no negated literals. Also, as shown by Theorem 3.4 the network structure does not alter causality, so we will assume a star network.

**Case A:** $\Phi = 1$

There is polynomial transformation of $\Phi$ to a minimal form $\Psi$, such that $\Psi = \Phi$ and $\Psi$ only contains the minterm clauses of $\Phi$. For example, if $\Phi = (A \wedge B) \vee (A \wedge B \wedge C) \vee (C \wedge D)$, $\Psi = (A \wedge B) \vee (C \wedge D)$. The transformation is polynomial, as $\Psi$ includes a clause $C_i$ of $\Phi$ only if $\nexists C_j$ that contains a subset of the literals of $C_i$. We will show that a variable $X_i$ is a functional cause of $\Phi$, iff $X_i \in C_i$, where $C_i$ a clause of $\Psi$ that evaluates to true under current assignment, and $x_i^0 = 1$.

First of all, if $X_i$ is not in $\Psi$, then $X_i$ is not a cause of $\Phi$, as there is no assignment that makes $X_i$ counterfactual for $\Psi$ and therefore for $\Phi$, as $\Psi = \Phi$. Therefore, $X_i$ cannot be a cause of $\Phi$.

If $X_i$ is in $\Psi$, but $\forall C_i \in \Psi$ that contain $X_i$, $C_i$ evaluates to false, then $X_i$ cannot be a functional cause, because of monotonicity since there is no negation. Any set $\vec{S}$ that makes $X_i$ counterfactual, has to contain the variables of $C_i$ whose initial assignment was 0, denote them with set $\vec{S}_c \subset \vec{S}$, and $\vec{S}' = \vec{S} \backslash \vec{S}_c$. Then $\Psi(\neg x_i^0, \neg \vec{s}_c^0, \neg \vec{s}'^0) = 0$ and $\Psi(x_i^0, \neg \vec{s}_c^0, \neg \vec{s}'^0) = 1$. $\Psi$ can be written as $\Psi = C_i \vee \Psi'$. This means that $\Psi'(\neg \vec{s}^0) = 0$. Therefore, $\Psi(x_i^0, \vec{s}_c^0, \neg \vec{s}'^0) = (C_i^0) \vee \Psi'(\neg \vec{s}'^0) = 0$, which means that $\Delta P_\Phi(\vec{S}') \neq 0$, so $X_i$ is not a cause.

Also, if $x_i^0 = 0$, because of monotonicity, it is not possible to switch a formula from 1 to 0, by flipping $X_i$ from 0 to 1.

Now, if $X_i \in C_i$, where $C_i$ a clause of $\Psi$ that evaluates to true under current assignment, and $x_i^0 = 1$, we select $\vec{S} = \{X_j \mid X_j \notin C_i \text{ and } x_j^0 = 1\}$. Then, if we write $\Psi = C_i \vee \Psi'$, we know that $\Psi'(\neg \vec{s}^0) = 0$, because $\Psi$ contains only minterms. That means that every clause $C_j$ has at least one variable that is not in $C_i$, and therefore can be negated by the above choice of $\vec{S}$. This makes $X_i$ counterfactual with contingency $\vec{S}$. Also $\Phi = \Psi$, therefore $\Delta P_\Phi = \Delta P_\Psi$. Since $X_i$ is counterfactual, $\Delta P_\Psi(X_i, \vec{S}) \neq 0$, and $\Delta P_\Psi(\vec{S}) = 0$. Also, $\vec{S}$ does not contain any variables of $C_i$, and therefore, for any subset of $\vec{S}$, $C_i$ is true, and therefore $\Psi$ is also true.

Therefore, $X_i$ is a cause of $\Phi$, iff $X_i \in C_i$, where $C_i$ a clause of $\Psi$ that evaluates to true under current assignment, and $x_i^0 = 1$, and this can be determined in polynomial time.

**Case B:** $\Phi = 0$

We again define the minterms transformation of $\Phi$ to $\Psi$, but this time by first eliminating the variables whose initial assignment is 1. For example, if $\Phi = (A \wedge B) \vee (C \wedge D) \vee (A \wedge E)$, with initial assignment $(a, b, c, d, e)^0 = (0, 1, 0, 1, 0)$, then $\Psi = A \vee D$. The clause $(A \wedge E)$ get eliminated because of the presence of minterm $A$. Similarly, we will show that $X_i$ is a cause of $\phi$, iff $X_i \in C_i$, where $C_i$ a clause of $\Psi$, and $x_i^0 = 0$.

First of all, if $X_i$ is not in $\Psi$, then $X_i$ is not a cause of $\Phi$, because either $x_i^0 = 1$, which eliminates it as a possible cause because of the monotonicity argument, or there was a minterm in $\Psi$ that caused its elimination. To make $X_i$ counterfactual in $\Phi$ through clause $C_i$, we need to invert all the variables $X_j \in C_i$ for which $x_j^0 = 0$ ($\vec{S}_c = \{X_j \mid X_j \in C_i \text{ and } x_j^0 = 0\}$). But because there is a minterm in $\Psi$, that contains a subset of these variables, the inversion would switch $C_j$ to true. $X_i$ will not be counterfactual unless we also invert at least one variable $X_k \in C_j$ for which $x_k^0 = 1$. So $\vec{S} = \vec{S}_c \cup \{X_k\}$. Then $\Delta P_\Phi(X_i, \vec{S}) \neq 0$, but for $\vec{S}' = \vec{S}_c$, $\Delta P_\Phi(\vec{S}') \neq 0$, which means that $X_i$ is not a cause.

If $X_i \in C_i$ and $C_i \in \Psi$, set $\vec{S} = \{X_j \mid X_j \in C_i \text{ and } x_j^0 = 0\}$. That makes $X_i$ counterfactual in $\Psi$ (and also in $\Phi$), as we know that there are no other clauses in $\Psi$ that contain a subset of $\vec{S}$ causing them to result to true. Also, obviously, for any subset of $\vec{S}$, $\Psi$ as well as $\Phi$ result to 0, which is the initial assignment.

Therefore, $X_i$ is a functional cause of $\Phi$, iff $X_i \in C_i$, where $C_i$ a clause of $\Psi$, and $x_i^0 = 0$, and this can be determined in polynomial time. □

Proof Theorem 3.14 (Positive CNF). Assume $\Phi$ a formula in CNF with no negated literals. Also, as shown by Theorem 3.4 the network structure does not alter causality, so we will assume a star network.

**Case A:** $\Phi = 1$

We define the maxterms transformation of $\Phi$ to $\Psi$, also eliminating variables with initial assignment 0. Due to monotonicity, any variable $X_i$ with $x_i^0 = 0$ cannot be a cause. As an example, if $\Phi = (A \vee B \vee C) \wedge (A \vee B) \wedge (A \vee D)$, and initial assignment $(a, b, c, d) = (1, 1, 1, 0)$, then $\Psi = A$. This is because the clause $(A \vee B \vee C)$ gets eliminated because of the presence of maxterm $(A \vee B)$, and because $D = 0$, $(A \vee B)$ also gets eliminated because of the creation of maxterm $A$. We will show that $X_i$ is a functional cause of $\Phi = 1$ iff $X_i \in \Psi$, which is computable in polynomial time.

If $X_i \notin \Psi$, then either $x_i^0 = 0$, in which case $X_i$ cannot be a cause, or $X_i$ was part of an eliminated clause $C_i$. $C_i = C_i' \vee X_i$ was eliminated because there was another clause $C_j \in \Phi$ which can be split into $C_j = C_j^+ \vee C_j^-$, so that $C_j^-$ evaluated to 0 under given assignment, and $C_j^+ \subseteq C_i'$. If $X_i$ is a cause in $\Phi$, then $\exists \vec{S}$, s.t. $\Delta P_\Phi(\neg X_i, \neg \vec{S}) \neq 0$ and $\forall \vec{S}' \subseteq \vec{S}, \Delta P_\Phi(\vec{S}) = 0$. There has to be $\vec{S}_c \subseteq S$, such that $\vec{S}_c = \{X_j \mid X_j \in C_i' \text{ and } x_j^0 = 1\}$, in other words, we need to set to 0 all variables in $C_i'$ in order to make $X_i$ counterfactual, and $C_i'$ has to contain at least one variable set to true, otherwise $X_i$ would not have been eliminated. Since $C_j^+ \subseteq C_i'$, then $C_j^+ \subseteq \vec{S}_c$. That means that inverting $\vec{S}_c$ would set $C_j$ to false. For $X_i$ to be counterfactual, $\vec{S}$ also needs to contain at least one variable from $C_j^-$, call it $X_c^-$. However, for $\vec{S}' = \vec{S} \setminus \{X_c^-\}$, $C_j$ would be set to false, so $\Delta P_\Phi(\vec{S}') \neq 0$, which means that $X_i$ cannot be a cause.

If $X_i \in C_i \in \Psi$ then set $\vec{S} = \{X_j \mid X_j \in C_i' \text{ and } x_j^0 = 1\}$, where $C_i = C_i' \vee X_i$. Inverting $\vec{S}$ does not invert $\Psi$ (or $\Phi$), as there is no $C_j \in \Phi$ or $\Psi$ that is $C_j \subseteq C_i'$, otherwise $C_i$ would have been eliminated since $C_j$ would have been a maxterm. Therefore $X_i$ is counterfactual with contingency $\vec{S}$. Also $\forall \vec{S}' \subseteq \vec{S} \; \Delta P_\Phi(\vec{S}') = 0$, as no clause can be negated if we invert fewer positive terms due to monotonicity. Therefore $X_i$ is a cause.

Since the transformation from $\Phi$ to $\Psi$ is polynomial, causality of $X_i$ can be determined in polynomial time.

**Case B:** $\Phi = 0$

From $\Phi$ we construct $\Psi$, which just contains the maxterms of $\Phi$. For example, if $\Phi = (A \vee B \vee C) \wedge (A \vee B) \wedge (A \vee D)$, then $\Psi = (A \vee B) \wedge (A \vee D)$. It is always $\Psi = \Phi$. We will show that $X_i$ is a functional cause of $\Phi$, iff $X_i \in C_i$, where $C_i$ a clause in $\Psi$ that evaluates to false.

Assume that $X_i$ is a cause of $\Phi$ in clause $C_i = C_i' \vee X_i$. Then $\exists \vec{S}$, such that $\Delta P_\Phi(\neg X_i, \neg \vec{S}) \neq 0$ and $\forall \vec{S}' \subseteq \vec{S}$, $\Delta P_\Phi(\vec{S}) = 0$. If $C_i'$ evaluates to 1 under given assignment, then $\vec{S}$ has to contain $\vec{S}_c = \{X_j \mid X_j \in C_i' \text{ and } x_j^0 = 1\}$. Also every clause $C_j \not\equiv C_i$, has to be set to true after the inversion of $\vec{S}$, otherwise $X_i$ would not be counterfactual. That means that $\vec{S}$ should also contain a subset $\vec{S}_R$ that sets all other clauses to 1. But then, for $\vec{S}' = \vec{S} \setminus S_c$, the formula will evaluate to 1, because $C_i'$ will evaluate to one along with all other clauses, which would make $X_i$ not a cause. Therefore, $C_i'$ evaluates to 0 under given assignment. Still, the fact that $\vec{S}$ contains a variable $X_j$ from every clause other than $C_i$ means that every clause contains a variable $X_j$ that is not contained in $C_i'$. Therefore, $\not\exists C_j$ such that $C_j \subseteq C_i'$, and therefore $C_i$ is a maxterm. Therefore, if $X_i$ is a functional cause of $\Phi$, then $X_i \in C_i$, where $C_i$ a clause in $\Psi$ that evaluates to false.

If $X_i \in C_i$, where $C_i$ a clause in $\Psi$ that evaluates to false, then set $\vec{S} = \{X_j \mid X_j \notin C_i \text{ and } x_j^0 = 0\}$. That will set all clauses apart from $C_i$ to true. That is guaranteed because $C_i$ is a maxterm, and therefore, $\not\exists C_j \subseteq C_i'$, where $C_i = C_i' \vee X_i$. Then $X_i$ is counterfactual for $\Psi$ and $\Phi$, and also, $\forall \vec{S}' \subseteq \vec{S} \; \Delta P_\Phi(\vec{S}') = 0$, as $C_i$ is stuck to false. Therefore, $X_i$ is a functional cause.

Since the maxterm transformation is polynomial, causa-

## D. DETAILS SECTION 4

## D.1 Complexity of aggregates

PROOF LEMMA 4.4 (SUM POSSIBLE CAUSES). Assume $\omega^0$ the result of a summation query and $\vec{X}^+$ the set of all the true tuples, and $\vec{X}^-$ the false ones.

**Case A: Q=WHY SO?, op="$\geq$":** $\phi = (\omega^0 > c) = 1$.

Assume $X_i \in X^-$. If $X_i$ is a cause of $\phi$, there must exist support $\vec{S} \subseteq \vec{X} \setminus \{X_i\}$ such that $\Delta P(X_i, \vec{S}) \neq 0$. Also $\vec{S}$ is partitioned into $\vec{S}^+ = \vec{S} \cap \vec{X}^+$ and $\vec{S}^- = \vec{S} \cap \vec{X}^-$. Assume $\omega_s, \omega_s^+, \omega_s^-$ the sum of values in $\vec{S}, \vec{S}^+$ and $\vec{S}^-$ respectively, and $\omega_t$ the value of $t$. Then $\omega^0 - \omega_s^+ + \omega_s^- + \omega_t < c \Rightarrow \omega_s^+ > \omega^0 - c + \omega_s^- + \omega_t > 0$. So, $\vec{S}^+ \neq \emptyset$.

Also, $\omega^0 - \omega_s^+ + \omega_s^- + \omega_t < c \Rightarrow \omega^0 - \omega_s^+ < c - \omega_t - \omega_s^- < c$ because all values are positive. Therefore, for $\vec{S}'_t = \vec{S}^+ \subseteq \vec{S}$, $\Delta P(\vec{S}'_t) \neq 0$, so a false tuple cannot be a cause of case A.

**Case B: Q=WHY SO?, op="$\geq$":** $\phi = (\omega^0 < c) = 1$.

Assume $X_i \in X^+$. If $X_i$ is a cause of $\phi$, there must exist support $\vec{S} \subseteq \vec{X} \setminus \{X_i\}$ such that $\Delta P(X_i, \vec{S}) \neq 0$. Also $\vec{S}$ is partitioned into $\vec{S}^+ = \vec{S} \cap \vec{X}^+$ and $\vec{S}^- = \vec{S} \cap \vec{X}^-$. Assume $\omega_s, \omega_s^+, \omega_s^-$ the sum of values in $\vec{S}, \vec{S}^+$ and $\vec{S}^-$ respectively, and $\omega_t$ the value of $t$. Then $\omega^0 - \omega_s^+ + \omega_s^- - \omega_t > c \Rightarrow \omega_s^- > c - \omega^0 - c + \omega_s^+ + \omega_t > 0$. So, $\vec{S}^- \neq \emptyset$.

Also, $\omega^0 - \omega_s^+ + \omega_s^- - \omega_t > c \Rightarrow \omega^0 + \omega_s^- > c + \omega_t + \omega_s^+ > c$ because all values are positive. Therefore, for $\vec{S}'_t = \vec{S}^- \subseteq \vec{S}$, $\Delta P(\vec{S}'_t) \neq 0$, so a true tuple cannot be a cause of case B.

**Case C: Q=WHY NO?, op="$\leq$":** $\phi = (\omega^0 < c) = 0$.

Assume $X_i \in X^-$. If $X_i$ is a cause of $\phi$, there must exist support $\vec{S} \subseteq \vec{X} \setminus \{X_i\}$ such that $\Delta P(X_i, \vec{S}) \neq 0$. Also $\vec{S}$ is partitioned into $\vec{S}^+ = \vec{S} \cap \vec{X}^+$ and $\vec{S}^- = \vec{S} \cap \vec{X}^-$. Assume $\omega_s, \omega_s^+, \omega_s^-$ the sum of values in $\vec{S}, \vec{S}^+$ and $\vec{S}^-$ respectively, and $\omega_t$ the value of $t$. Then $\omega^0 - \omega_s^+ + \omega_s^- + \omega_t < c \Rightarrow \omega_s^+ > \omega^0 - c + \omega_s^- + \omega_t > 0$. So, $\vec{S}^+ \neq \emptyset$.

Also, $\omega^0 - \omega_s^+ + \omega_s^- + \omega_t < c \Rightarrow \omega^0 - \omega_s^+ < c - \omega_t - \omega_s^- < c$ because all values are positive. Therefore, for $\vec{S}'_t = \vec{S}^+ \subseteq \vec{S}$, $\Delta P(\vec{S}'_t) \neq 0$, so a false tuple cannot be a cause of case C.

**Case C: Q=WHY NO?, op="$\geq$":** $\phi = (\omega^0 > c) = 0$.

Assume $X_i \in X^+$. If $X_i$ is a cause of $\phi$, there must exist support $\vec{S} \subseteq \vec{X} \setminus \{X_i\}$ such that $\Delta P(X_i, \vec{S}) \neq 0$. Also $\vec{S}$ is partitioned into $\vec{S}^+ = \vec{S} \cap \vec{X}^+$ and $\vec{S}^- = \vec{S} \cap \vec{X}^-$. Assume $\omega_s, \omega_s^+, \omega_s^-$ the sum of values in $\vec{S}, \vec{S}^+$ and $\vec{S}^-$ respectively, and $\omega_t$ the value of $t$. Then $\omega^0 - \omega_s^+ + \omega_s^- - \omega_t > c \Rightarrow \omega_s^- > c - \omega^0 - c + \omega_s^+ + \omega_t > 0$. So, $\vec{S}^- \neq \emptyset$.

Also, $\omega^0 - \omega_s^+ + \omega_s^- - \omega_t > c \Rightarrow \omega^0 + \omega_s^- > c + \omega_t + \omega_s^+ > c$ because all values are positive. Therefore, for $\vec{S}'_t = \vec{S}^- \subseteq \vec{S}$, $\Delta P(\vec{S}'_t) \neq 0$, so a true tuple cannot be a cause of case D. $\square$

PROOF PROP. 4.5 (WHY SO? = WHY NO?). The premise of this statement is straightforward: If $\omega^0$ the SUM value, and $t$ is a cause of $\phi = (\omega^0 \geq c) =$ true. Define condition $\psi = (\omega^0 < c) = (\omega^0 \not\geq c)$. Clearly $\phi \Rightarrow \neg \psi$, so if $t$ is a cause of $\phi$ it is also a cause of $\neg \psi$.

Similarly, if $t$ is a cause of $\psi$ $(\omega^0 < c)=$true, then $\psi \Rightarrow \neg \phi$, and therefore $t$ is also a cause of SUM$\not\geq c$. $\square$

PROOF THEOREM 4.6 (SUM HARDNESS). We will use a reduction from the *subset sum problem* (SSP): given $n$ positive numbers and a target bound $c$, find a subset of the numbers summing to $c$. More formally: Let $\vec{V} = [v_1, \ldots, v_n]$ with

```
Initialize all K(0,j) = 0 and all K(d,0) = ∞
for j = 1 to n
    for d = 1 to ω⁰ − c + v
        if d < v_j:  K(d,j) = K(d,j−1)
        else:        K(d,j) = min[K(d − v_j, j−1) + 1, K(d,j−1)]
return min[K(ω⁰ − c + 1), ..., K(ω⁰ − c + v)]
```

**Figure 16:** Pseudo-polynomial time algorithm to determine causes for sum over one input relation in $\mathcal{O}(n(\omega^0 - c + v))$.

$v_i \in \mathbb{N}^+$ be a given vector of positive integers, $c$ a positive integer, and define $\Omega(\vec{X})$ as the dot product $\Omega(\vec{X}) = \vec{X}\vec{V}$, with $\vec{X} = [x_1, \ldots, x_n]$ and $x_i \in \{0, 1\}, 1 \leq i < n$ represents a vector of binary variables. The subset sum problem is to find an assignment of binary values $\vec{x}^0$ so that $\Omega(\vec{x}^0) = c$.

We reduce the above SSP problem to the following WHY SO? problem. Construct an ordered set of tuples $\vec{T}'$ with one attribute corresponding to the values of the vector $\vec{V}'$ with $v'_i = v_i, i \in \{1, \ldots, n\}$ and $v'_{n+1} = 1$. Now consider the aggregate SUM$(\vec{x}'^0)$ for actual assignment $\vec{x}'^0$ with $x'^0_i = 1$ for $i \in \{1, \ldots, n+1\}$. Then $t'_{n+1}$ is a WHY SO? explanation for the aggregate condition $\bigl(\text{SUM}(\vec{x}'^0) \geq c+1\bigr) = $ true iff there is one assignment $\vec{x}'^1$ with $x'^1_{n+1}$ for which SUM$(\vec{x}'^1) = c$.

Hence, we have reduced SSP to determining causality of tuples for the SUM aggregate. $\square$

PROOF THEOREM 4.7 (SUM PSEUDO-PTIME). Determining responsibility of a tuple $t$ with value $v$ for a WHY SO? aggregate condition $\bigl(\text{SUM}(\vec{x}^0) \geq c\bigr) = $ true is solvable in pseudo-polynomial time $\mathcal{O}(nc)$ using the following dynamic programming algorithm.

Consider the new set $\vec{V}^*$ that consists of all values of $\vec{V}$ that are true under current assignment except for the value $v$. Let $\omega^* = \text{SUM}(\vec{V}^*)$. Then we have to find a minimal subset $\vec{V}'^* \subseteq \vec{V}^*$ whose values add up to a value in the closed interval $[\omega^* - c + 1, \omega^* - c + v]$. Now define the subproblem

$$K(d, j)$$

as the minimum subset size $|\vec{V}''^*|$ with values summing up to $d$ for the subset of values $\{v_1, \ldots, v_j\}$. We then express $K(d, j)$ in a way that either value $v_j$ is needed to achieve the minimal value, or it isn't needed:

$$K(d, j) = \min[K(d - v_j, j - 1) + 1, K(d, j - 1)]$$

The answer we seek is the minimal value of $\{K(d, n) \mid \omega^0 - c + 1 \leq d \leq \omega^0 - c + v\}$. The algorithm then consists of filling out a two-dimensional table, with $n$ rows and $\omega^0 - c + v + 1$ columns, hence in $\mathcal{O}(n(\omega^0 - c + v))$ time (Fig. 16). $\square$